\documentstyle[11pt]{article}
\newcommand{\be}{\begin{equation}}
\newcommand{\ee}{\end{equation}}
\newcommand{\bea}{\begin{array}}
\newcommand{\eea}{\end{array}}

\title{TWO KINDS OF DUALITY}
\author{Robert Carroll\\
University of Illinois\\
Urbana, IL 61801\\email:  rcarroll@symcom.math.uiuc.edu}

\date{February, 1997}

\begin{document}
\bibliographystyle{plain}
\maketitle

\begin{abstract} 
In a finite zone KdV context we show relations between the duality 
variables of Faraggi-Matone and those involved in Seiberg-Witten 
type duality.
\end{abstract}


\section{INTRODUCTION}
\renewcommand{\theequation}{1.\arabic{equation}}\setcounter{equation}{0}

We will indicate some relations between the duality
variables $X$ and $\psi$ in 
\cite{fa} and the variables of Seiberg-Witten (SW) type duality occuring in
$N=2$ susy Yang-Mills (YM) theory. 
A priori there is no relation between the variables, nor any relation to
KdV, but we show that if one takes in advance
a finite zone
KdV context on a hyperelliptic curve $\Sigma_g$ 
of genus $g$ then there are relations between the variables.
The mechanism involves looking 
at connections between an extended idea of
dispersionless KdV ($dKdV_{\epsilon}$ from \cite{cb,ce}) and the 
Whitham type theory related to $\Sigma_g$ based on \cite{cc,cd,cf,
da,ia,ka,na} for example.  An important ingredient here is indicated
in \cite{aa,cd,na} where the algebraic asymptotics based on vertex
operators at $P_{\infty}$ is related to the theta function formulas
for Baker-Akhiezer (BA) functions $\psi$.  No a priori conncection
to quantum mechanics is assumed.

\section{BACKGROUND ON RIEMANN SURFACES}
\renewcommand{\theequation}{2.\arabic{equation}}\setcounter{equation}{0}

We recall first some ideas on BA functions and Riemann surfaces following
\cite{aa,cc,cd,ka,kb,na}.  Given a compact Riemann surface $\Sigma_g$
of genus $g$ let $(A_i,B_i)$ be a canonical homology basis, $d\omega_j$ a 
basis of normalized holomorphic differentials ($\oint_{A_j}d\omega_i=
\delta_{ij}$), ${\cal A}(P)=(\int_{P_0}^Pd\omega_k)$ the Abel-Jacobi
map ($P_o\not= P_{\infty}\sim\infty$), and $\Theta(z)=\Theta[0](z)$
the Riemann theta function.  Let $\lambda^{-1}$ be a local coordinate
near $\infty$ with $\lambda(P_{\infty})=\infty$ and take $d\Omega_j
=d(\lambda^j+O(\lambda^{-1}))$ to be normalized meromorphic differentials
of the second kind ($\oint_{A_j}d\Omega_i=0$).  Other normalizations are
also used (e.g. $\Re\oint_{A_i}d\Omega_j=\Re\oint_{B_i}d\Omega_j=0$)
but we will not dwell on this.  We set also $\Omega_{jk}=\oint_{B_k}
d\Omega_j$.  Now let $D=P_1+\cdots+P_g$ be a nonspecial divisor of degree
$g$ and set $z_0=-K-{\cal A}(D)$ where $K\sim (K_j)$ corresponds to
Riemann constants.  One can now introduce ``time" coordinates $t_j$ via
a uniquely defined BA function (up to normalization)
\be
\psi = exp(\int^P_{P_0}\sum t_nd\Omega_n)
\cdot\frac{\Theta({\cal A}(P) + \sum (t_j/2\pi i)(\Omega_{jk})+z_0)}
{\theta({\cal A}(P)+z_0)}
\label{A}
\ee
(see \cite{cc,cd} for an extensive discussion - we are working here in
in a KP framework for convenience).  Next one defines a dual divisor
$D^*$ via $D+D^*-2P_{\infty}\sim K_{\Sigma}$ where $K_{\Sigma}$
is the canonical class of $\Sigma_g$ (class of meromorphic differentials).
Then the dual BA function is (up to normalization)
\be
\psi^*\sim e^{-\int_{P_o}^P\sum t_nd\Omega_n}\cdot \frac{\Theta
({\cal A}(P)-\sum (t_j/2\pi i)(\Omega_{jk})+z_0^*)}{\Theta({\cal A}(P)
+z_0^*)}
\label{B}
\ee
($z_0^*=-{\cal A}(D^*)-K$)
and the BA conjugate differential is $(\clubsuit\clubsuit\clubsuit)
\,\,\psi^{\dagger}=
\psi^*d\hat{\Omega}$ where ($E\sim$ prime form)
\be
d\hat{\Omega}(P')=\frac{\Theta({\cal A}(P')+z_0)\Theta({\cal A}(P')+z_0^*)}
{E(P,P_{\infty})^2}
\label{C}
\ee
Thus $d\hat{\Omega}$ has zero divisor $D+D^*$ and a unique double pole
at $P_{\infty}$ so that $\psi\psi^*d\hat{\Omega}=\psi\psi^{\dagger}$
is meromorphic with a second order pole at $P_{\infty}$ and no other poles.
Note here in (\ref{A}) for example there should be a normalization factor
$c(t)$ multiplying the right side (cf. \cite{db}); we will incorporate
the normalizations via theta functions in the calculations below.
\\[3mm]\indent
It is instructive and useful to enlarge the context in the spirit of
\cite{cc,cg,cn,ia,na}.  We stay in a KP framework and write
(normalizations are now included)
\be
\psi=exp\left[\sum_1^{\infty}t_j\left(\int_{P_0}^Pd\Omega^j+\Omega^j(P_0)
\right)+i\sum_1^g
\alpha_j\left(\int_{P_0}^Pd\omega_j+\omega_j(P_0)\right)\right]\times
\label{D}
\ee
$$\times\frac{\Theta\left({\cal A}(P)+\sum_1^{\infty}(t_j/
2\pi i)(\Omega_{jk})
+i\sum_1^g\alpha_j(B_{jk})+z_0\right)\Theta({\cal A}(P_{\infty})
+z_0)}{\Theta\left({\cal A}(P_{\infty})+\sum_1^{\infty}
(t_j/2\pi i)(\Omega_{jk})
+i\sum_1^g\alpha_j(B_{jk})+z_0\right)\Theta({\cal A}(P)+z_0)}$$
(note $\int^Pd\Omega_j\sim\int_{P_0}^Pd\Omega_j
+\Omega_j(P_0)$) and ${\cal A}(P)=(\int_{P_{\infty}}^Pd\omega_j)+{\cal A}
(P_{\infty})$) and explicitly now ($z=\lambda^{-1}$ amd $q_{mj}=q_{jm}$)
\be
d\Omega_j=d\Omega^j\sim d\left(\lambda^j-\sum_1^{\infty}\frac{q_{mj}}{m}
z^m\right);\,\,d\omega_j\sim d\left(-\sum_1^{\infty}
\sigma_{jm}\frac{z^m}{m}\right);\,\,\Omega_{nj}=2\pi i\sigma_{jn}
\label{E}
\ee
(see \cite{cc} for details).  There is also a general theory of
prepotential etc. following \cite{cc,ia,na} for example which involves
\be
d{\cal S}=\sum_1^ga_jd\omega_j+\sum_1^{\infty}T_nd\Omega_n;\,\,
\frac{\partial d{\cal S}}{\partial a_j}=d\omega_j;\,\,\frac{\partial d{\cal S}}
{\partial T_n}=d\Omega_n
\label{F}
\ee
If we consider functions $F(a,T)$ related to $d{\cal S}$ via
\be
\frac{\partial F}{\partial a_j}=\frac{1}{2\pi i}\oint_{B_j}d{\cal S};\,\,
\partial_nF=-Res_{\infty}z^{-n}d{\cal S}
\label{G}
\ee
then, given the standard 
class of solutions of the Whitham hierarchy satisfying (cf. \cite{cc,kc})
\be
2F=\sum_1^ga_j\frac{\partial F}{\partial a_j}+\sum_1^{\infty}T_n\frac
{\partial F}{\partial T_n}
\label{H}
\ee
there results
\be
2F=\sum_1^g\frac{a_j}{2\pi i}\oint_{B_j}d{\cal S}-\sum_1^{\infty}
T_nRes_{\infty}z^{-n}d{\cal S}
\label{I}
\ee
Writing now, in the notation of \cite{na},
$d\omega_j=-\sum_1^{\infty}\sigma_{jm}z^{m-1}dz$ with $d\Omega_n=
[-nz^{-n-1}-\sum_1^{\infty}q_{mn}z^{m-1}]dz$, and using (\ref{F}),
one obtains ($B_{jk}$ is the period matrix)
\be
2F=\frac{1}{2\pi i}\sum_{j,k=1}^gB_{jk}a_ja_k+2\sum_1^ga_j\sum_1^{\infty}
\sigma_{jk}T_k+\sum_{k,l=1}^{\infty}q_{kl}T_kT_l
\label{J}
\ee
\indent
Thus the expression (\ref{J}) comes from the Riemann surface theory, 
without explicit reference to the BA function, and we
consider now (\ref{D}) and 
\be
\psi=exp\left(\sum_1^{\infty}t_i\lambda^i\right)\times\frac{\tau
(t-[\lambda^{-1}],\alpha)}{\tau(t,\alpha)}
\label{K}
\ee
to which ideas of dKP can be applied to 
introduce the slow variables $T_k$.  This means that we will
be able to introduce
slow variables in two different ways and the resulting comparisons will show
an equivalence of procedures.  In practice this will enable one to treat
$\epsilon$ on the same footing in the Whitham theory and in the dispersionless
theory
(see also \cite{cd} for an approach based on \cite{aa}).
Thus from (\ref{D}) and (\ref{K}) one obtains 
an expression for $\tau$ of the form
($t_1=x,\,\,t_2=y,\,\,t_3=t,\,\cdots$)
\be
\tau(t,\alpha)=exp[\hat{F}(\alpha,t)]\Theta\left(
{\cal A}(P_{\infty})+\sum_1^{\infty}(t_j/2\pi i)(\Omega_{jk})
+i\sum_1^g\alpha_j(B_{jk})+z_0\right)
\label{L}
\ee
where $k=1,\cdots,g$ and
\be
\hat{F}(\alpha,t)=\frac{1}{2}\sum_{k,l=1}^{\infty}q_{kl}t_kt_l-\frac
{1}{4\pi i}\sum_{j,k=1}^{\infty}B_{jk}\alpha_j\alpha_k+i\sum_1^g
\alpha_j\sum_1^{\infty}\sigma_{jk}t_k+\sum_1^{\infty}d_kt_k
\label{M}
\ee
(see also \cite{kd} for a similar form - recall here ${\cal A}(P)=
(\int_{P_0}^Pd\omega_j)$ and $P_0\not= P_{\infty}$ is required).
Putting in the slow variables $T_k=\epsilon t_k$ and $a_k=i\epsilon\alpha_k$
one will find
that the quadratic part of $\hat{F}(T/\epsilon,a/i\epsilon)$ 
in $T$ and $a$ is exactly $F(a,T)/\epsilon^2$ for $F$ in (\ref{J});
here $\tau=exp[(1/\epsilon^2)\tilde{F}+O(1/\epsilon)]$
(with $\tilde{F}/\epsilon^2$ the quadratic part of $\hat{F}(T/\epsilon,
a/i\epsilon)$)
is the natural form of $\tau$ based
on (\ref{K}) and it is associated with $\psi\sim exp[(1/\epsilon)
S+O(1)]$ (cf. \cite{ci} - note this is $S$ and not ${\cal S}$ - $S$
will be discussed later in Section 4).
In \cite{na} one writes then from (\ref{L}) and (\ref{D})
respectively 
\be
\frac{1}{\epsilon^2}F(a,T)+O\left(\frac{1}{\epsilon}\right)=
log\tau\left(\frac{T}{\epsilon},\frac{a}{i\epsilon}\right)=\epsilon^{-2}
\sum_0^{\infty}\epsilon^nF^{(n)}(T,a);
\label{N}
\ee
$$dlog\psi\left(p,\frac{T}{\epsilon},
\frac{a}{i\epsilon}\right)=\epsilon^{-1}\sum_0^{\infty}\epsilon^n
d{\cal S}^{(n)}(p,T,a)\sim \frac{1}{\epsilon}S+O(1)$$
where $d{\cal S}^{(0)}\sim d{\cal S}$ in (\ref{F}) and $F^{(0)}\sim F$
in (\ref{J}).  Suitable calculations are displayed in \cite{cc} to
establish the relations between $F$ and $\hat{F}$ as indicated.
\\[3mm]\indent
For perspective however let us make now a few background observations.  First
we refer first to \cite{ch} where it
is proved that $F_{mn}=F_{nm}$ in ${\cal B}_n=\lambda^n-\sum_1^{\infty}
(F_{nm}/m)\lambda^{-m}$ (the $F_{mn}$ being treated as algebraic symbols
with two indices generally and $F_{mn}=\partial_m\partial_nF$
specifically).
Since near the point at infinity we have $\Omega_n\sim
\lambda^n-\sum_1^{\infty}(q_{mn}/m)\lambda^{-m}$ the same sort of proof
by residues is suggested ($F_{mn}=-Res_{\lambda}[{\cal B}_nd\lambda^m]$)
but we recall that ${\cal B}_n=\lambda^n_{+}$ so there is an underlying
$\lambda$ for all ${\cal B}_n$ which makes the proof possible.  Here 
one should be careful however.  For example 
$(\spadesuit\spadesuit\spadesuit)\,\,
p=\lambda-\sum_1^{\infty}(H_j/j)\lambda^{-j}$ corresponds to $P=\lambda
+\sum_1^{\infty}P_{j+1}\lambda^{-j}$ in \cite{ch} with $P_{j+1}=F_{1j}/j$
(i.e. $H_j\sim -F_{1j}$) and the ``inverse" is $\lambda=P+\sum_1^{\infty}
U_{n+1}P^{-n}$ (arising from a Lax operator $L$ via dKP).  The corresponding
inverse for $(\spadesuit\spadesuit\spadesuit)$ 
then characterizes $\lambda$ in terms
of $p$ but one does not automatically 
expect $\Omega_n\sim\lambda^n_{+}$.  The matter is somewhat
subtle.  Indeed the BA function is defined from the Riemann surface via
$d\Omega_n,\,\,d\omega_j$, and normalizations.  It then produces a unique
asymptotic expansion at $\infty$ which characterizes $\psi$ near $\infty$ in
terms of $\lambda$ and hence must characterize the $d\Omega_n$ and $d
\omega_j$ asymptotically.  Moreover the normalizations must be built into
these expansions since they were used in determining $\psi$.  
Thus we must have $F_{mn}\sim q_{mn}$ as a consequence of the
BA function linking the differentials and the asymptotic expansions
(note also that the 
formal algebraic determination of ${\cal B}_n$ via $\lambda^n_{+}$
is a consequence of relating the $d\Omega_n$ to operators $L_n=
L^n_{+}$ as in \cite{ka} which corresponds to looking at $\lambda^n_{+}$
with $\lambda=P+\sum_1^{\infty}U_{n+1}P^{-n}$ as above). 
Another approach
(following \cite{cd}) is to extract from remarks after (\ref{M}) that
$q_{mn}=F_{mn}$ at $T^0_k=0$ via $F_{mn}=\partial_m\partial_nF$,
so that expanding around an arbitrary
$T_k^0$ as in \cite{kc} one can assert that $q_{mn}=F_{mn}$ with 
arbitrary argument.  Further
with this identification we recover the Whitham equations as in \cite{cd}
via
\be
\partial_k\Omega_n=-\sum_{m,n=1}^{\infty}\frac{F_{mnk}}{m}z^m=
\partial_n\Omega_k=-\sum_{m,n=1}^{\infty}\frac{F_{mkn}}{m}z^m
\label{O}
\ee
Finally we recall now that in SW duality one sets $a_j^D=\partial F/
\partial a_j$ and the formulas (\ref{F}) - (\ref{I}) are fundamental
relations (see e.g. \cite{cc,cg,da,ia,ka,na}).  This theme is the first
kind of duality in consideration here.

\section{BACKGROUND ON $(X,\psi)$ DUALITY}
\renewcommand{\theequation}{3.\arabic{equation}}\setcounter{equation}{0}

We extract first from \cite{ce} to indicate the duality of \cite{fa}
between $X$ and $\psi$ (cf. also \cite{ba,ca,cb}).
The point of departure is the Schr\"odinger equation
\be
{\cal H}\psi_E=-\frac{\hbar^2}{2m}\psi''_E+V(X)\psi_E=E\psi_E
\label{AA}
\ee
where $X$ is the quantum mechanical (QM) space variable with $\psi'_E=
\partial\psi_E/\partial X$ and we write $\epsilon=\hbar/\sqrt{2m}\,\,
(E$ is assumed real).  In \cite{ca,cb} we discussed the possible origin of
this from a Kadomtsev-Petviashvili (KP) situation $L^2_{+}\psi=\partial\psi/
\partial t_2$ where $L^2_{+}=\partial_x^2-v(x,t_i)$ and e.g. $\tau_2=
-i\sqrt{2m}T_2$ so $\partial_{t_2}=\epsilon\partial_{T_2}=-i\hbar
\partial_{\tau_2}$ (one writes $X=\epsilon x$ and 
$T_i=\epsilon t_i$ in the dispersionless theory). 
As seen below the format of dispersionless theory is related to WKB methods
and in fact we will expand the standard dispersionless theory in the WKB
direction. 
This leads to an
approximation
\be
\epsilon^2\psi''_E-V(X,T_i)\psi_E\sim\epsilon\frac{\partial\psi_E}{\partial T_2}
=-i\hbar\frac{\psi_E}{\partial\tau_2} 
\label{BB}
\ee
corresponding to the Schr\"odinger equation.  This is also related to the
Korteweg-deVries (KdV) equation and it's dispersionless form dKdV as
indicated below.  
For the approximation one
assumes e.g. $v=v(x,t_i)\to v(X/\epsilon,T_i/\epsilon)
=V(X,T_i)
+O(\epsilon)$.  This is
standard in dispersionless KP = dKP and certainly realizable
by quotients of homogeneous polynomials for example. 
In fact it is hardly a restriction since given e.g. $F(X)=\sum_0^{\infty}
a_nX^n$ consider $\tilde{f}(x,t_i)=a_0+\sum_1^{\infty}(x^n/\prod_2^{n+1}
t_i)$.  Then $\tilde{f}(X/\epsilon,T_i/\epsilon)=a_0+\sum_1^{\infty}
(X^n/\prod_2^{n+1}T_i)$ and one can choose the $T_i$ recursively so that
$1/T_1=a_1,\,\,1/T_1T_2=a_2,\cdots$, leading to $F(X)=\tilde{F}(X,T_i)$. 
Further, when
$\psi_E=exp(S/\epsilon)$ for example, one has $\epsilon\psi'_E=S_X\psi_E$
with $\epsilon^2\psi''_E=\epsilon S_{XX}\psi_E+(S_X)^2\psi_E$ so in (\ref{BB})
we are neglecting an $O(\epsilon)\psi_E$ term from $v$, and for $\psi_E
=exp(S/\epsilon)$ another $\epsilon S_{XX}\psi_E$ term is normally removed
in dispersionless theory.  Then for ${\cal H}$ independent of $\tau_2$ for
example one could assume $V$ is independent of $T_2$ and write formally
in (\ref{BB}), $\hat{\psi}_E=exp(E\tau_2/i\hbar)\cdot\psi_E$, 
with ${\cal H}\psi_E
=E\psi_E$, which is (\ref{AA}).  Since in the QM problem one does not
however run $\hbar\to 0$ (hence $\epsilon\not\to 0$) one should argue
that these $O(\epsilon)$ terms should be retained,
and we will develop this approach, which essentially corresponds to WKB
(with some background structure).
In particular one could ask
for $v(X/\epsilon,T_i/\epsilon)=V(X,T_i)+\epsilon\hat{V}
(X,T_i)+O(\epsilon^2)$ and retain the 
$\epsilon\hat{V}$ term along with $\epsilon
S_{XX}$, in requiring e.g. $S_{XX}=\hat{V}$ (this is covered 
below - an additional term also arises).
In fact the passage from $v\to V$ or $V+\epsilon \hat{V}$ is the only
``assumption" in our development and this admits various realizations;
the impact here only involves some possible minor restrictions on
the class of quantum potentials
to which the theory applies.  The background mathematics behind
$V$ determined by KP or KdV essentially generates some additional structure
which allows us to insert $X$ into the theory in a manner commensurate
with its role in \cite{fa}.  The formulation of \cite{fa} then entails
some constraints on the background objects as indicated in the text.
We emphasize that inserting $S$ is familiar from WKB (cf. \cite{mb,mc});
we are introducing in an ad hoc manner additional
variables $T_i$ or $T_i,\,\,\lambda$ or $k$, etc. to spawn a KP or KdV
theory.  We do not assume or even suggest that this is in any way connected
a priori with the physics of the quantum mechanical problem (although
of course it conceivably could be since integrability ideas are important
in quantum mechanics).  This procedure generates a nice Hamilton-Jacobi (HJ)
theory which guides one to insert $X$ into the machinery, but the
insertion itself is at ``ground level" and simply reflects a WKB
formulation; neither the underlying KdV or KP dynamics nor the HJ
theory is directly used here.  Once $X$ is involved connections to
\cite{fa} are immediate.  Actually the procedure could be reversed
as a way of introducing duality ideas into the $\epsilon$-dispersionless
theory of \cite{cb,ce}) and this should probably be
related to the duality already studied in Whitham theory (cf. 
\cite{cc,cg,ia,ka,ma}), given a finite zone theory on a Riemann surface.
Thus start with KdV or KP, go to the Schr\"odinger
equation and $dKdV_{\epsilon}$ or $dKP_{\epsilon}$, develop the HJ theory,
and then use \cite{fa} to create duality.  More generally, start from
a finite zone KdV situation with associated Whitham dynamics on a
Riemann surface and compare dualities; this is the aim of the
present paper.
\\[3mm]\indent
We list first a few of the equations from \cite{fa}, as written in 
\cite{ca,cb,ce}, without a discussion
of philosophy (some of which will
be mentioned later).  Thus ${\cal F}$ is a prepotential
and, since $E$ is real,
$\psi_E$ and $\bar{\psi}_E=\psi_E^D$ both satisfy (\ref{AA}) with $\psi_E^D=
\partial{\cal F}/\partial\psi_E$.  
The Wronskian in (\ref{AA}) is taken to be
$W=\psi'\bar{\psi}-\psi\bar{\psi}'=2\sqrt{2m}/i\hbar=2/i\epsilon$ and one
has ($\psi=\psi(X)$ and $X=X(\psi)$ with $X_{\psi}=\partial X/\partial\psi=
1/\psi'$)
\be
{\cal F}'=\psi'\bar{\psi};\,\,{\cal F}=\frac{1}{2}\psi\bar{\psi}+\frac
{X}{i\epsilon};\,\,
\frac{\partial\bar{\psi}}{\partial\psi}=\frac{1}{\psi}\left[\bar{\psi}-
\frac{2}{i\epsilon}X_{\psi}\right]
\label{FF}
\ee
($\psi$ always means $\psi_E$ but we omit the subscript occasionally
for brevity).
Setting $\phi=\partial{\cal F}/\partial(\psi^2)=\bar{\psi}/2\psi$ with
$\partial_{\psi}=2\psi\partial/\partial(\psi^2)$ and evidently
$\partial\phi/\partial
\psi=-(\bar{\psi}/2\psi^2)+(1/2\psi)(\partial\bar{\psi}/\partial\psi)$
one has a Legendre transform pair
\be
-\frac{X}{i\epsilon}=\psi^2\frac{\partial{\cal F}}{\partial(\psi^2)}
-{\cal F};\,\,-{\cal F}=\phi\frac{1}{i\epsilon}X_{\phi}-\frac{X}{i\epsilon}
\label{HH}
\ee
One obtains also $(\spadesuit\bullet\spadesuit)\,\,
|\psi|^2=2{\cal F}-(2X/i\epsilon)\,\,({\cal F}_{\psi}=\bar{\psi});\,\,
-(1/i\epsilon)X_{\phi}=\psi^2;\,\,{\cal F}_{\psi\psi}=
\partial\bar{\psi}/\partial\psi$.
Further from $X_{\psi}\psi'=1$ one has $X_{\psi\psi}\psi'+X^2_{\psi}\psi''
=0$ which implies
\be
X_{\psi\psi}=-\frac{\psi''}{(\psi')^3}=\frac{1}{\epsilon^2}\frac
{(E-V)\psi}{(\psi')^3}
\label{KK}
\ee
\be
{\cal F}_{\psi\psi\psi}=\frac{E-V}{4}({\cal F}_{\psi}-\psi\partial^2_{\psi}
{\cal F})^3=\frac{E-V}{4}\left(\frac{2X_{\psi}}{i\epsilon}\right)^3
\label{LL}
\ee
Although a direct comparison of (\ref{LL}) to the Gelfand-Dickey resolvant
equation ($(\clubsuit\clubsuit)$ below) 
is not evident ($V'$ is lacking) a result of
T. Montroy which expands ${\cal F}_{\psi\psi\psi}$
shows that in fact (\ref{LL}) corresponds exactly to
\be
\epsilon^2{\cal F}'''+4(E-V)\left({\cal F}'-\frac{1}{i\epsilon}\right)
-2V'\left({\cal F}-\frac{X}{i\epsilon}\right)=0
\label{ZZZ}
\ee
which is $(\clubsuit\clubsuit)$ since $\Xi=|\psi|^2=2{\cal F}-(2X/i\epsilon)$.
\\[3mm]\indent
Next there is a so-called eikonal transformation (cf. \cite{mf}) which can
be related to \cite{fa} as in \cite{ca,cb,ce}.  We consider
real $A$ and $S$ with
$\psi=Ae^{(i/\hbar)S};\,\,p=ASin[(1/\hbar)S];\,\,q=ACos[(1/\hbar)
S]$.  Then
introducing new variables 
$\chi=A^2=|\psi|^2;\,\,\xi=(1/2\hbar)S$
it follows that there will be a Hamiltonian format with symplectic form
$(\spadesuit)\,\,\delta p\wedge
\delta q=\delta\xi\wedge\delta\chi=\tilde{\omega}$.
It is interesting to write down the connection between the $(S,A)$ or
$(\chi,\xi)$ type
variables and the variables from \cite{fa} and it will be useful to  
take now $\psi=Aexp(iS/\epsilon)\,\,(\epsilon=\hbar/\sqrt{2m})$ with
$\xi\sim S/2\epsilon$. Then
\be
{\cal F}=\frac{1}{2}\chi+\frac{X}{i\epsilon};\,\,{\cal F}'=
\psi'\bar{\psi} =\frac{1}{2}\chi'
+\frac{i}{\epsilon}P\chi
\label{TT}
\ee
for $S'=S_X=P$ and there is an interesting relation
$(\clubsuit)\,\,
P\chi=-1\Rightarrow \delta\chi=-(\chi/P)\delta P$.
Further from $\phi=(1/2)exp[-(2i/\epsilon)S]$ and $\psi^2=\chi exp(4i\xi)$ 
we have
\be
\psi^2\phi=\frac{1}{2}\chi=-\frac{1}{\epsilon}\phi X_{\phi};\,\,\xi=\frac
{S}{2\epsilon}=\frac{i}{4}log(2\phi)
\label{VV}
\ee
Now the theory of the Seiberg-Witten (SW) differential $\lambda_{SW}$ 
following \cite{ca,cc,cg,da,ia,ka,ma} for example 
involves finding a differential
$\lambda_{SW}$ of the form $QdE$ or $td\omega_0$ (in the spirit of
\cite{ka} or \cite{da,ia} respectively) such that $d\lambda_{SW}=\omega$
is a symplectic form.  In the
present context one can ask now whether the form $\tilde{\omega}$ of 
$(\spadesuit)$ makes any sense in such a context.  Evidently this is jumping
the gun since there is no Riemann surface in sight (see however \cite{ca}
for a Riemann surface with some validation as in \cite{cc,cd}); 
the motivation to consider 
the matter here comes from the
following formulas which express $\tilde{\omega}$ nicely in terms of
the duality variables of \cite{fa} (another version of a ``canonical"
symplectic form in terms of ${\cal F}$ alone is given below).
Thus a priori $\psi=\Re\psi+i\Im\psi$ has two components which are also
visible in $\psi=Aexp(iS/\epsilon)$ as $A$ and $S$.  The relation $P\chi
=\chi(\partial S/\partial X)=-1$ indicates a dependence between $A$
and $S'$ (but not $A$ and $S$) which is a consequence of the duality between
$\psi$ and $X$.  Then $2AS'\delta A+A^2\delta S'=0$ or $\delta S'
=-(2S'/A)\delta A\equiv (\delta S'/S')=-2(\delta A/A)$, whereas $\delta
\psi/\psi\sim 2(\delta A/A) +(i/\epsilon)\delta S$.  It follows that
$\Re(\delta\psi/\psi)=-(\delta S'/S')$ and $\Im(\delta \psi/\psi)=(\delta
S/\epsilon)$.  The sensible thing seems to be to look at the complex
dependence of $X(\psi)$ and $\psi(X)$ in terms of two real variables
and $\delta\xi\wedge\delta\chi$ will have a nice
form in transforming to the variables of \cite{fa}.  In particular
from $\psi^2\phi=(1/2)\chi$ with $\delta\chi=4\phi\psi\delta\psi+
2\psi^2\delta\phi$ we obtain
$(\delta\psi/\psi)=2(\delta\chi/\chi)-
(\delta\phi/\phi)$.
Hence one can write 
\be
\delta\xi\wedge\delta\chi=\frac{i}{4}\frac{\delta\phi}{\phi}
\wedge\chi\frac{\delta\chi}{\chi}=\frac{i}{2}\delta\phi\wedge\delta\psi^2
\sim\frac{i}{2}\delta\bar{\psi}\wedge\delta\psi
\label{XX}
\ee
(note $\delta\phi=(1/2\phi)\delta\bar{\psi}-(\bar{\psi}/2\psi^2)\delta\psi$)
and in an exploratory spirit the differentials $\lambda=(i/2)\phi\delta
\psi^2$ or $\lambda=(i/2)\psi^2\delta\phi$, along
with $\lambda=(i/2)\bar{\psi}\delta\psi$
or $\lambda=(i/2)\psi\delta\bar{\psi}$, might merit further consideration.
\\[3mm]\indent
We refer now to \cite{ch,ci,cj,ta} for dispersionless KP ($=$ dKP) and consider
here $\psi=exp[(1/\epsilon)S(X,T,\lambda)]$ instead of $\psi=Aexp(S/\epsilon)$
(more details are given later).
Thus $P=S'=S_X$ and $P^2=V-E$ 
but $E\not= \pm\lambda^2$ (unless otherwise stated) and
this does not define $S$ via $P=S_X$ unless we have a KdV situation
(which does not
seem a priori desirable but in fact seems to be the natural format here
upon development with modifications
of the dispersionless theory);  
thus generally $\lambda$ is the $\lambda$ of $S(T_n,\lambda)$ from KP theory
and we recall that $\psi$ always means $\psi_E$ as in \cite{fa}.
Some routine calculation yields
(recall $X_{\psi}=1/\psi'$ and $\psi'=(P/\epsilon)\psi$)
\be
\phi=\frac{1}{2}e^{-(2i/\epsilon)\Im S};\,\,\frac{1}{\epsilon}X_{\phi}=
-ie^{(2/\epsilon)S};\,\,X_{\psi}=\frac{\epsilon}{P}e^{-S/\epsilon}
\label{AAQ}
\ee
\be
\frac{1}{\epsilon}X_{\psi\psi}=\frac{E-V}{P^3}e^{-S/\epsilon};\,\,
{\cal F}_{\psi}=\bar{\psi}=e^{\bar{S}/\epsilon};\,\,
{\cal F}_{\psi\psi}=e^{-(2i/\epsilon)\Im S}-\frac{2}{iP}e^{-2S/\epsilon}
\label{ABQ}
\ee
\be
|\psi|^2=e^{(2/\epsilon)\Re S};\,\,\frac{S}{\epsilon}=\frac{1}{2}log|\psi|^2
-\frac{1}{2}log(2\phi);\,\,\bar{P}=\bar{S}_X=P-\frac{2}{i\psi\bar{\psi}}
\label{ACQ}
\ee
Summarizing one has
\be
\Im{\cal F}=-\frac{X}{\epsilon};\,\,\Re{\cal F}=\frac{1}{2}|\psi|^2=
\frac{1}{2}e^{\frac{2}{\epsilon}\Re S}=-\frac{1}{2\Im P}
\label{AGQ}
\ee
In the present situation $|\psi|^2=exp[(2/\epsilon)\Re S]$ and $2\phi=
exp[-(2i/\epsilon)\Im S]$ can play the roles of independent variables
(cf. (\ref{ACQ}).  The version here of $P\chi=-1$ is $\chi\Im P=-1$,
while $\psi^2\phi=(1/2)|\psi|^2=(1/2)\chi$ again, and
we obtain as above the formula (\ref{XX}). 
Now note that for $L=\partial+\sum_1^{\infty}u_i
\partial^{-i},\,\,L^2_{+}=\partial^2+2u_1$, and $u_1=\partial^2log(\tau)$
where $\tau$ is the famous tau function.
This implies $v=-2\partial^2log(\tau)$ here, from which $V=-2F_{XX}$ for
$\tau=exp[(1/\epsilon^2)F+O(1/\epsilon)]$ 
in the dispersionless theory (cf. \ref{N})).  We recall also 
the Gelfand - Dickey
resolvant equation (cf. \cite{ce}) for $\Xi=\psi\bar{\psi}$, namely,
in the present notation 
$(\clubsuit\clubsuit)\,\,
\epsilon^2\Xi'''-4V\Xi'-2V'\Xi+4E\Xi'=0$
(direct calculation).  
Using $\Xi=
2{\cal F}-(2X/i\epsilon),\,\,\Xi'=2{\cal F}'-(2/i\epsilon),\,\,\Xi''=
2{\cal F}''$, and $\Xi'''=2{\cal F}'''$, we obtain 
then from $(\clubsuit\clubsuit)$ (cf. also (\ref{ZZZ}))
\be
\epsilon^2{\cal F}'''+\left({\cal F}'-\frac{1}{i\epsilon}\right)(8F''
+4E)+4F'''\left({\cal F}-\frac{X}{i\epsilon}\right)=0
\label{AKQ}
\ee
which provides a relation between $F$ and ${\cal F}$.
We will see in Section 3 how to embellish all this with a new modification of
the dKP and dKdV theory.
Thus we state here
\\[3mm]\indent {\bf THEOREM 3.1.}$\,\,$  Under the hypotheses indicated
the equation (\ref{AKQ}) yields a relation between the prepotential
${\cal F}$ of $(X,\psi)$ duality defined via (\ref{FF}) and the prepotential
$F(a,T)$ of (\ref{J}) (also corresponding to a free energy in dKP or dKdV).
\\[3mm]\indent {\bf REMARK 3.2.}$\,\,$  We emphasize that the development
here is first order in WKB and heuristic; the exposition to follow using an
expanded $dKdV_{\epsilon}$ theory based on \cite{cb} will establish
precise relations.
\\[3mm]\indent
One sees that the Riemann surface background produces the $a_i$ variables
naturally here and we want now to find a definition of ${\cal F}$ which is
based on dKdV quantities and not on $\psi$ directly.  Perhaps this will
suggest another way to view duality based on ${\cal F}$.  One notes that
the word duality involving ${\cal F}$ refers to $X$ and $\psi$ whereas 
duality in SW theory refers to $a_i$ and $a_i^D=\partial F/\partial a_i$
as being dual.  In ${\cal F}$ of (\ref{FF}) of course $\bar{\psi}=\psi^D=
\partial {\cal F}/\partial\psi$ but it is $X$ and $\psi$ which are said to
be dual.  It will be shown in Section 3 (following \cite{cb,ce}) that
$dX\wedge dP\sim [-\epsilon/2(\Re{\cal F})^2]d\Im({\cal F})\wedge
d(\Re{\cal F})$ follows from the WKB aspects
of dKP where $P=S_X$.  On the othe hand, following \cite{ca,cc,ka}, one has
a canonical symplectic form $\omega\sim\sum da_i\wedge d\omega_i$ associated
with SW theory.  A priori there seems to be no conceptual reason why SW
theory should have any relation to $(X,\psi)$ duality, except perhaps 
that the background mathematics and development in \cite{ba,ma} has many
features related to SW mathematics.  The connection indicated by (\ref{AKQ})
relating $F(a,T)$ and ${\cal F}$ is momentarily purely formal; it may
not signify much in terms of conceptual meaning and this will be
pursued below.  We note also the natural occurance of a symplectic form
$(i/2)\delta\bar{\psi}\wedge \delta\psi$ in (\ref{XX}) whose ``duality"
analogue would seem to involve $\sum da_i^D\wedge da_i$; there seems to
be no immediate conceptual connection here however.

\section{DISPERSIONLESS THEORY}
\renewcommand{\theequation}{4.\arabic{equation}}\setcounter{equation}{0}

We give next a brief sketch of some ideas regarding dispersionless KP
(dKP) following mainly \cite{ch,ci,cj,ke} to which we refer for 
philosophy (cf. also \cite{mb,mc} for WKB). 
We will make various notational adjustments as we go along 
and subsequently will modify some of the theory.  One
can think of fast and slow variables with $\epsilon x=X$ and $\epsilon t_n=
T_n$ so that $\partial_n\to\epsilon\partial/\partial T_n$ and $u(x,t_n)
\to\tilde{u}(X,T_n)$ to obtain from the KP equation $(1/4)u_{xxx}+3uu_x
+(3/4)\partial^{-1}\partial^2_2u=0$ the equation $\partial_T\tilde{u}=3
\tilde{u}\partial_X\tilde{u}+(3/4)\partial^{-1}(\partial^2\tilde{u}/
\partial T_2^2)$ when $\epsilon\to 0$ ($\partial^{-1}\to(1/\epsilon)
\partial^{-1}$).  For the underlying mathematical theory write
$(t_n)$ for
$(x,t_n)$ (i.e. $x\sim t_1$ here) and consider
$L_{\epsilon}=\epsilon\partial+\sum_1^{\infty} u_{n+1}(\epsilon,T)
(\epsilon\partial)^{-n}$.
Here $L$ is the Lax operator and one takes now
$\psi=exp[(1/
\epsilon)S(T,\lambda)+O(1)];\,\,
\tau=exp[(1/\epsilon^2)F(T)+O(1/\epsilon)]$ (we suppress $a$ in $F(a,T)$
for simplicity).
Recall that $\partial_nL=[B_n,L],\,\,B_n=L^n_{+},
\,\,L\psi=\lambda\psi,$
and $\psi=\tau(T-(1/n\lambda^n))exp[\sum_1^{\infty}T_n\lambda^n]/
\tau(T)$.  Putting in the $\epsilon$ and using $\partial_n$ for
$\partial/\partial T_n$ now, with $P=S_X$, one obtains
$(\bullet)\,\,
\lambda=P+\sum_1^{\infty}U_{n+1}P^{-n};\,\,
P=\lambda-\sum_1^{\infty}P_i\lambda^{-1};
\,\,\partial_nS={\cal B}_n(P)$ which implies $\partial_nP=\hat{\partial}
{\cal B}_n(P)$.
Further $\hat{\partial}\sim \partial_X+(\partial P/\partial X)\partial_P$ and
$B_n=\sum_0^nb_{nm}\partial^m$ so one has ${\cal B}_n=\sum_0^nb_{nm}
P^m$.  
We list a few additional formulas which are 
easily obtained (cf. \cite{ci}); thus, writing $\{A,B\}=\partial_PA\partial A
-\partial A\partial_PB$ one has
$\partial_n\lambda=\{{\cal B}_n,\lambda\}$ and 
we can write $S=\sum_1^{\infty}T_n\lambda^n+\sum_1^{\infty}S_{j+1}
\lambda^{-j}$ with $S_{n+1}=-(\partial_nF/n),\,\,
\partial_mS_{j+1}=-(F_{mn}/n),
\,\,V_{n+1}=-nS_{n+1}$,
with
\be
{\cal B}_n=\lambda^n+\sum_1^{\infty}\partial_nS_{j+1}\lambda^{-j};\,\,
\partial S_{n+1}\sim -P_n\sim -\frac{\partial V_{n+1}}{n}\sim
-\frac{\partial\partial_n F}{n}
\label{YE}
\ee
\indent
We sketch next a few formulas from \cite{ch,ci,ke}.  First
it will be convenient to rescale the $T_n$ variables and write 
$t'=nt_n,\,\,T_n'=nT_n,\,\,
\partial_n=n\partial'_n=n(\partial/\partial T'_n)$.  Then
$\partial'_nS=\lambda^n_{+}/n;\,\,\partial'_n\lambda=\{{\cal Q}_n,
\lambda\}\,\,({\cal Q}_n={\cal B}_n/n);$ etc.
Now think of $(P,X,T'_n),\,\,n\geq 2,$ as basic Hamiltonian variables
with $P=P(X,T'_n)$.  Then $-{\cal Q}_n(P,X,T'_n)$ will serve as a
Hamiltonian via $(\spadesuit\spadesuit)\,\,
\dot{P}'_n=dP'/dT'_n=\partial{\cal Q}_n;\,\,\dot{X}'_n=
dX/dT'_n=-\partial_P{\cal Q}_n$.
The function
$S(\lambda,X,T_n)$ plays the role of part of a generating function 
$\hat{S}$ for the Hamilton-Jacobi (HJ) theory with action angle variables 
$(\lambda,-\xi)$ where
$d\lambda/dT'_n=\dot{\lambda}'_n=\partial_{\xi}R_n=0;\,\,
d\xi/dT'_n=\dot{\xi}'_n=-\partial_{\lambda}R_n=-\lambda^{n-1}$
(note that $\dot{\lambda}'_n=0\sim\partial'_n\lambda=\{{\cal Q}_n,
\lambda\}$).  
The motivation here for HJ theory is to provide a guide for inserting
$X$ from the dispersionless
context into the framework in fundamental manner, commensurate with its
role in \cite{fa} (cf. also \cite{ci,cm}).
For KdV one 
looks at the dispersionless theory based on $k$ where $\lambda^2
\sim(ik)^2=-k^2$.  There results, for $P=S_X$, the 
formula $P^2+q=-k^2$, and we write
${\cal P}=(1/2)P^2+p=(1/2)(ik)^2$ with $q\sim 2p\sim 2u_2$.  One has
$\partial k/\partial T_{2n}=\{(ik)^{2n},k\}=0$ and from $ik=P(1+qP^{-2})^
{1/2}$ we obtain
\be
ik=P\left(1+\sum_1^{\infty}{\frac{1}{2}\choose m}q^mP^{-2m}\right);\,\,
P=ik-\sum_1^{\infty}
P_j(ik)^{-j}
\label{YR}
\ee
(cf. $(\bullet)$ with $u_2=q/2$).  The flow equations become then
$\partial'_{2n+1}P=\hat{\partial}{\cal Q}_{2n+1};\,\,\partial'_{2n+1}(ik)
=\{{\cal Q}_{2n+1},ik\}$.
Note here some rescaling is needed since we want $(\partial^2+q)^{3/2}_{+}=
\partial^3+(3/2)q\partial+(3/4)q_x=B_3$ instead of our previous $B_3\sim
4\partial^3 +6q\partial+3q_x$.  Thus we want ${\cal Q}_3=(1/3)P^3+(1/2)qP$
to fit the notation above.  
\\[3mm]\indent
Now
the dKP theory as in \cite{ch,ci,ke,ta} involves a parameter $\epsilon\to
0$ and we recall $L=\partial+\sum_1^{\infty}u_{n+1}(t)\partial^{-n}\to
L_{\epsilon}=\epsilon\partial +\sum_1^{\infty}u_{n+1}(\epsilon,T)(\epsilon
\partial)^{-n}$ where $t\sim (t_k),\,\,T\sim (T_k),$ and $X=T_1$ with
$u_{n+1}(\epsilon,T)=U_{n+1}(T)+O(\epsilon)$ as in Section 2.  Then for
$\psi=exp(S/\epsilon)$ one has $L\psi=\lambda\psi\to\lambda= P+\sum_1^{\infty}
U_{n+1}P^{-n}$ where $P=S_X$ with $S=S(X,T_k,\lambda)\,\,(k\geq 2)$.
Here all the terms which are $O(\epsilon)$ are passed to zero and in
view of $\epsilon\not\to 0$ in the QM situation where $\epsilon=\hbar/\sqrt
{2m}$ one thinks of rewriting some of the dKP theory in order to retain
$O(\epsilon)$ terms at least (and dropping $O(\epsilon^2)$ terms).  
We will call this $dKP_{\epsilon}$ theory and it essentially corresponds
to an expanded WKB with the proviso 
that there is a background mathematics
providing some additional structure. 
We recall now 
$S=\sum_1^{\infty}T_n\lambda^n+\sum_1^{\infty}S_{j+1}\lambda^{-j};\,\,
{\cal B}_n=\partial_nS=\lambda^n+\sum_1^{\infty}\partial_nS_{j+1}
\lambda^{-j}$
and via $log\psi=(S/\epsilon)+O(1)\sim log\tau[\epsilon,T_n-(\epsilon/
n\lambda^n)]-log\tau+\sum_1^{\infty}T_n\lambda^n/\epsilon$ with $log\tau
=(F/\epsilon^2)+O(1/\epsilon)$ and $F[T_n-(\epsilon/n\lambda^n)]-F(T_n)\sim
-\epsilon\sum_1^{\infty}(\partial_nF/n\lambda^n)+O(\epsilon^2)$ one 
obtains $S_{n+1}=-(\partial_nF/n)$.  Consider now the next order terms
via $F$ (recall the $a_i$ variables are suppressed here - cf. \cite{cc})
\be
F\left(T_n-\frac{\epsilon}{n\lambda^n}\right)-F(T_n)=-\epsilon\sum_1^{\infty}
\left(\frac{\partial_nF}{n\lambda^n}\right)+\frac{\epsilon^2}{2}\sum
\left(\frac{F_{mn}}{mn}\right)\lambda^{-m-n}+O(\epsilon^3)
\label{five}
\ee
Thus $\Delta log\tau=(1/\epsilon^2)\Delta F$ has $O(1)$ terms $(1/2)\sum
(F_{mn}/mn)\lambda^{-m-n}$ which correspond to the $O(1)$ terms in $log\psi$.
Hence we have a natural way of writing
$\tilde{S}=S^0+\epsilon S^1$ with $S^0=S$ and
\be
S^1=\frac{1}{2}\sum\left(\frac{F_{mn}}{nm}\right)\lambda^{-m-n};\,\,
\tilde{S}_X\sim P+\frac{\epsilon}{2}\sum\left(\frac{F_{1mn}}{nm}\right)\lambda^
{-m-n}
\label{six}
\ee
One could also include $F=F^0+\epsilon F^1$, etc. 
(as in (\ref{N})) with e.g. $\hat{V}=P_X
+2PP^1=-2F^1_{XX}$ but we restrict matters here to $F=F^0$ and $\hat{V}
=0$ (it will be seen below that $F^1=0$ is appropriate).
\\[3mm]\indent
It turns out that dKP, dKdV, and $dKP_{\epsilon}$
will not do (cf. \cite{cb,ce}) and we sketch a few points.  
Thus note first that
the equation ${\cal F}=(1/2)\psi\bar{\psi}+(X/i\epsilon)\sim (1/2)exp
[(2/\epsilon)\Re S] +(X/i\epsilon)$ has $\epsilon$ at various levels
which is confusing.  Moreover $|\psi|^2=exp[(2/\epsilon)\Re S]$ should
be bounded by $1$ which suggests a $dKP_{\epsilon}$ format
with $S\to \tilde{S}=S^0+\epsilon S^1\,\,(S^0\sim S),\,\, \Re S^0=0$, and
\be
|\psi|^2=e^{2\Re S^1}=exp\left[\Re\sum\left(\frac{F^0_{mn}}{mn}\right)
\lambda^{-m-n}\right]
\label{OF}
\ee
For this to occur we need $(\bullet\bullet\bullet)\,\,
\Re S^0=\Re\sum_1^{\infty}T_n\lambda^n+\Re\sum_1^{\infty}S_{j+1}\lambda^
{-j}=0$
where one expects $S_{j+1}=-(\partial_jF^0/j)$ to be real.  This suggests
that it would be productive to think of KdV after all with $\lambda=ik$
imaginary, $T_{2n}=0,$ and $\partial_{2n}F^0=0$ as indicated below (so
$S_{2n+1}=0$ and only $\lambda^{-j}$ terms occur in 
$(\bullet\bullet\bullet)$ for $j$
odd).  
For $dKdV_{\epsilon}$  
one establishes $F^0_{m,2n}=0$ as in \cite{ch} (cf. below)
so in (\ref{OF}) one only has terms $(\bullet\heartsuit\bullet)\,\,
(F^0_{(2m+1)(2n+1)}/(2n+1)(2m+1))\cdot\lambda^{-2(m+n)-2}$
which would be real for $\lambda=ik$.  Thus $S^0$ and $P=S^0_X$ are
imaginary while $S^1$ and $P^1=\partial_X S^1$ are real. 
The conditions under which the formulas of 
\cite{fa} are valid with $E=\pm\lambda^2$ real involve $\lambda$ either
real or pure imaginary.  
Thus a KdV situation is indicated with $\lambda
=ik,\,\,\lambda^2=-k^2=-E$ but we will 
need $dKdV_{\epsilon}$.
To see this note for dKdV we
will have $P$ purely imaginary with $U_j$ and $P_j$ real
and only odd powers of $P$ or 
$k$ appear in (\ref{YR}).  Look now at 
$ik=P(1+\sum_1^{\infty}U_mP^{-2m})$ and for $P=iQ$ we see
that $(ik)^{2n+1}_{+}={\cal B}_{2n+1}$ will be purely imaginary.  Further
$\partial_P{\cal B}_{2n+1}$ will involve only even powers of $P$ and hence
will be real.  
Thus write now 
\be
{\cal B}_{2n+1}=\sum_0^nb_{nj}P^{2j+1};\,\,\partial_P{\cal B}_{2n+1}=
\sum_0^n(2j+1)b_{nj}P^{2j}
\label{LM}
\ee
and we have $(\clubsuit\bullet\clubsuit)\,\,
(d/dT_n)\Im{\cal F}=-(1/\epsilon)\dot{X}_n=(1/\epsilon)
\sum_0^n(2j+1)b_{nj}P^{2j}$.
Then the condition $P=iQ$ leads to a compatible KdV situation
$(\clubsuit\bullet\clubsuit)$ and further
$\dot{P}_n=dP/dT_n=\partial{\cal B}_n=\sum_0^n\partial (b_{nj})
P^{2j+1}$
which is realistic (and imaginary).
Now we note that there
is danger here of a situation where $\Re P=0$ implies $\Re S=0$
which in turn would imply $|\psi|^2=1$ (going against the philosophy
of keeping $|\psi|^2$ as a fundamental variable) and this is one reason
we will need $dKdV_{\epsilon}$ with (\ref{OF}) - $(\bullet\heartsuit\bullet)$,
etc.  Thus in general $(\bullet\bullet)\,\,
S=\sum_1^{\infty}T_n\lambda^n-\sum_1^{\infty}(\partial_n F/n)\lambda^{-n}$
and
$P=\lambda-\sum_1^{\infty}(F_{1n}/n)\lambda^{-n}\,\,
(F\sim F^0$ here) while ${\cal B}_m
=\lambda^n-\sum_1^{\infty}(F_{mn}/n)\lambda^{-n}$ 
and for KdV (with $\lambda=ik$) it follows from the residue formula
(cf. \cite{ch}) that
$F_{nm}=F_{mn}=Res_P\left(\lambda^md\lambda^n_{+}\right)$
that $F_{m,2n}=0$ and from a $\bar{\partial}$ analysis (cf. \cite{ch,ci})
$\partial_jF=(j/2i\pi)\int\int \zeta^{j-1}\bar{\partial}_{\zeta}S
d\zeta\wedge d\bar{\zeta}$.
The $\partial_jF$ and $F_{1j}$ can be computed explicitly as in \cite{ch}
and in particular $F_{1,2n}=0$ with ($P^2-U=-k^2$)
$F_{1,2n-1}=(-1)^n\left(U/2\right)^n\prod_1^n[(2j-1)/j]$.
A further calculation along the same lines also shows that $F_{2n}=\partial_
{2n}F=0$ for KdV.
Generally $F$ will be real along with the $F_{mn}$ and we recall that the
expression for ${\cal B}_{2m+1}$ arising from $(\bullet\bullet)$
is an alternate
way of writing (\ref{LM}).  For $\lambda=ik,\,\,P$ and ${\cal B}_{2m+1}$
will be purely imaginary but $S$ could be complex 
via $\sum_1^{\infty}T_n\lambda^n$ since all powers $\lambda^n
=(ik)^n$ will occur in $(\bullet\bullet)$.  Thus $\Re S\not=0$ and we have a
perfectly respectable situation, provided the $T_{2n}$ are real.
However $T_{2n}$ imaginary as in KP1 (cf. \cite{cb}), or 
as in (\ref{B}), would imply 
$\Re S=0$ and $|\psi|^2=1$ which is not desirable.
Another problem is that if $\Re S\not= 0$ is achieved via the times then
$|\psi|^2\sim exp[(1/\epsilon)\sum T_{2n}\lambda^{2n}]$ will not necessarily
be $\leq 1$.  This and other arguments rejecting $dKP_{\epsilon}$ lead
one now to $dKdV_{\epsilon}$ as the natural framework (cf. \cite{cb}
for details).
\\[3mm]\indent
Now for $dKdV_{\epsilon}$,
in view of (\ref{OF}) - $(\bullet\bullet\bullet)$, etc., 
there is no problem with $\Re S^0=0$
while happily 
$\Re S^1\not= 0$ and $|\psi|^2\leq 1$ is realistic.  The equation
(\ref{YR}) applies now but we cannot write
$ik\sim\tilde{P}(1+q\tilde{P}^{-2})^{\frac{1}{2}}$ for $\tilde{P}=
P+\epsilon P^1$.
Indeed other terms will arise involving $P_X$ for example since,
for 
$\tilde{S}=S^0+\epsilon S^1$ with $\psi=exp(\tilde{S}/\epsilon)=exp
[(S^0/\epsilon)+S^1]$, we have $\tilde{P}=\partial\tilde{S}=S^0_X+
\epsilon S^1_X=P+\epsilon P^1$ so that $\epsilon\partial\psi=\tilde{P}
\psi=(P+\epsilon P^1)\psi,\,\,\epsilon^2\partial^2\psi=(\epsilon P_X
+\epsilon^2P_X^1)\psi +2\epsilon P^1P\psi+P^2\psi+\epsilon^2 (P^1)^2\psi$,
etc. 
along with $\epsilon\partial(\psi/\tilde{P})=
-\epsilon(\tilde{P}_X/\tilde{P}^2)\psi+\psi=\psi
-\epsilon((P_X/P^2)\psi+O(\epsilon^2)$
from which $(\epsilon\partial)\psi\to\psi$ or $(\epsilon\partial)^{-1}\psi
\to \psi/\tilde{P}$ in some sense.  Continuing such calculations we obtain
terms of $O(\epsilon)$ in $(1/\psi)(\epsilon\partial)^{-n}\psi$ of the form
$(1/\psi)
{n+1 \choose 2}(\epsilon\partial)^{-n}\left(P_X\psi/P^2\right)$
and from $L\psi=\lambda\psi$ we get to first order
$\lambda=P+\sum_1^{\infty}U_{n+1}P^{-n}+O(\epsilon)$ with a 
complicated $O(\epsilon)$ term
(see (\ref{FU}) for some clarification of this).
Note here also that $P
=ik-\sum_1^{\infty}P_n(ik)^{-n}$ inverts (\ref{YR}) with $P_n=0$ for
$n$ even ($P_n=F_{1n}/n$ here - cf. \cite{ch} where there is an index
shift in the $P_n$); this shows that $P=iQ$.  
Further the constraint $|\psi|^2\Im\tilde{P}=-1\equiv |\psi|^2\Im P
=-1$ and this can be written $exp(2S^1)\Im P=-1$. 
In any event this leads to expressions for $ik,\,\,(ik)_{+}^{2n+1}$, etc.
and in particular 
for $P=iQ$ imaginary and
$S^1,\,P^1$ real we obtain
$2S^1+log(\Im P)=i\pi\Rightarrow 2P^1=-(\Im P_X/\Im P)\Rightarrow
P_X=-2PP^1$.
We do not pursue this here however since
in fact the HJ theory is not crucial
here as far as $\tilde{P}=P+\epsilon P^1$ is concerned.  Given $S=S^0
+\epsilon S^1$ and $F=F^0$ we know $\tilde{P}=P+\epsilon P^1$
is correct and that is all that is needed for the formulas of \cite{fa}.
Further calculations suggest that one can obtain exact balances 
for the HJ theory (perhaps
with constraints) but higher powers of $\epsilon$ should be included
(cf. also \cite{mb,mc}); in fact the development in Section 5 should suffice
for this but we do not pursue the matter here.
\\[3mm]\indent
Thus we take $\lambda^2=-E$ and specify $dKdV_{\epsilon}$.  We can still
label $\psi$ as $\psi_E$ but now one can imagine a $T_2\sim\tau$ variable
inserted e.g. via $\psi=\psi(X,T_{2n+1})exp(E\tau/i\hbar)\,\,
(n\geq 0)$ with $i\hbar\psi_{\tau}=E\psi$ and $\epsilon^2\psi''-V\psi
=-E\psi=\lambda^2\psi$ where $V=V(X,T_{2n+1})$ etc.
Consider ${\cal F}=(1/2)\psi\bar{\psi}+(X/i\epsilon)$ with $\psi=
exp[(1/\epsilon)S^0+S^1],\,\,\Re S^0=0$ as in $(\bullet\bullet\bullet)$,
and $|\psi|^2=exp(2\Re S^1)$ as in (\ref{OF}).  Here
\be
S^0=i\left[\sum_1^{\infty}T_{2n+1}(-1)^nk^{2n+1}+
\sum_1^{\infty}S_{2n}(-1)^nk^{-2n+1}\right]
\label{OK}
\ee
and explicitly
\be
{\cal F}=\frac{1}{2}exp\left[\sum_1^{\infty}\left(\frac{F^0_{(2m+1)(2n+1)}}
{(2m+1)(2n+1)}\right)(-1)^{m+n+1}k^{-2(m+n+1)}\right]+\frac{X}{i\epsilon}
\label{OL}
\ee
Thus the $\epsilon$ ``problem" has been removed from the $|\psi|^2$ term but
$\epsilon$ still occurs as a scale factor with $X$.
Look now at (\ref{ACQ}) with $P$ replaced by $\tilde{P}$ to obtain
$|\psi|^2\Im\tilde{P}=-1$ which in view of the $\epsilon$ independence
of $|\psi|^2$ suggests that $\Im P^1=0$ which in fact is true from 
$(\bullet\heartsuit\bullet)$.
Thus $|\psi|^2\Im P=-1$ as before but $P=S^0_X$ now.  Next for $\phi=
\bar{\psi}/2\psi$ we have $\phi=(1/2)exp[-(2i/\epsilon)\Im S]$ and 
$S^0$ is imaginary as in (\ref{OK}) with $S^1$ real as indicated in 
$(\bullet\heartsuit\bullet)$.
Consequently
\be
\phi=\frac{1}{2}exp\left[-\frac{2i}{\epsilon}\Im S^0\right]=
\label{OM}
\ee
$$=\frac{1}{2}exp\left[-\frac{2i}{\epsilon}\left(\sum_0^{\infty} 
(-1)^nT_{2n+1}k^{2n+1}+\sum_1^{\infty}(-1)^nS_{2n}k^{-2n+1}\right)\right]$$
One can also return to the discussion at the end of Section 4.1 and,
in the same heuristic first order spirit,
suggest again that $X=-\epsilon\Im {\cal F}$ and (for $P=iQ$)
\be
P=i\Im\tilde{P}=iQ=-\frac{i}{|\psi|^2}=-\frac{i}{2\Re
{\cal F}}
\label{ON}
\ee
are fundamental variables.  Note also from (\ref{OK}), $log(2\phi)=-(2/\epsilon)
S^0$, so
\be
log(2\phi)=-4i\xi=-\frac{2i}{\epsilon}\left(\sum (-1)^nT_{2n+1}k^{2n+1}+
\sum (-1)^nS_{2n}k^{-2n+1}\right)
\label{OOO}
\ee
This leads to a result from \cite{cb}, namely
\\[3mm]\indent {\bf THEOREM 4.1.}$\,\,$
From $dX\wedge dP$ there is a possibly fundamental symplectic form
\be
dX\wedge dP=-\epsilon d(\Im{\cal F})\wedge\frac{i}{2(\Re{\cal F})^2}
d(\Re {\cal F})=-\frac{i\epsilon}{2(\Re{\cal F})^2}d(\Im{\cal F})\wedge
d(\Re{\cal F})
\label{OP}
\ee
which 
seems intrinsically
related to the duality idea based on ${\cal F}$.
Note that this is not $dX\wedge d\tilde{P}$ (which would involve an additional
term $dX\wedge dP^1$, where a relation to $dX\wedge dP$ could then be 
envisioned via $P^1=-(1/2)\partial_Xlog\,P$).  In particular (\ref{OP})
is based only on first order WKB structure and is not dependent on KdV
connections (cf. Section 5 for expansion).
\\[3mm]\indent {\bf REMARK 4.2.}$\,\,$ We will refer to this as a ``naive"
theorem since an improved version arises in Section 5 from our expanded
theory.
\\[3mm]\indent  
The constraint $|\psi|^2\Im P=-1$ becomes
$exp[2\Re S^1]\Im S^0_X=-1$ which can be written out in
terms of $F^0=F$ and $\partial_XS_{2n}$ (cf. \cite{ca}).
Let us also compute the form 
$\omega =\delta\xi\wedge\delta\chi$ from (\ref{XX}) in
one of its many forms.  First recall $S^0$ is imaginary and $S^1$ is real
with $log(2\phi)=-(2/\epsilon)S^0=-4i\xi$ and $\chi=|\psi|^2=exp(2S^1)$.
Therefore formally, via $\xi=-(i/2\epsilon)S^0$, we have
$\omega=\delta\xi\wedge\delta\chi= -\left(i\chi/\epsilon\right)
\delta S^0\wedge\delta S^1$.
The difference here from (\ref{OP}) for example is that the term $X=-\epsilon
{\cal F}$ has no relation to $S^0$ or $S^1$ a priori.

\section{SYNTHESIS}
\renewcommand{\theequation}{5.\arabic{equation}}\setcounter{equation}{0}

Let us organize what we have so far.
From Section 2 we take a finite zone KdV
situation and produce a prepotential $F$ as in (\ref{J}) with asymptotic
connections to a BA function $\psi=exp[(1/\epsilon)S+O(1)]$ as in 
(\ref{N}) (where $\tau=exp[(1/\epsilon^2)F+O(1/\epsilon)]$ is also spelled
out).  Further one can make connections via the asymptotics of $\psi$ 
between $\Omega_n$ and ${\cal B}_n$ via $F_{mn}=q_{mn}$.  This brings
the $a_i$ variables into $F$ (and $d{\cal S}$) with ($F\sim F^0$)
\be
\partial_nF=F_n=-Res\,z^{-n}d{\cal S}=Res\,\lambda^ndS
\label{EA}
\ee
Note from $S=\sum_1^{\infty}T_n\lambda^n-\sum_1^{\infty}(\partial_jF/j)
\lambda^{-j}$ one has $dS=(\sum_1^{\infty}nT_n\lambda^{n-1}+\sum_1^{\infty}
\partial_nF\lambda^{-n-1})d\lambda$ whereas in Section 2 one is dealing
with
\be
-\partial_mF_n=F_{mn}=Res\,z^{-n}\partial_md{\cal S}=Res\,z^{-n}d\Omega_m
=-q_{mn}
\label{EB}
\ee
corresponding to $F_n=-Res\,z^{-n}d{\cal S}=F_n$.  Actually it is
interesting to compare the form of $dS$ with $d{\cal S}$ via
\be
d{\cal S}=\sum a_jd\omega_j+\sum T_nd\Omega_n=-\sum a_j(\sum\sigma_{jm}
z^{m-1})dz +
\label{ED}
\ee
$$+\sum T_n\left(n\lambda^{n-1}d\lambda-\sum q_{mn}z^{m-1}dz\right)=
\sum nT_n\lambda^{n-1}d\lambda-$$
$$-\sum z^{m-1}\left(\sum a_j\sigma_{jm}+
\sum T_nq_{mn}\right)dz$$
while $dS=\sum nT_n\lambda^{n-1}d\lambda-\sum\partial_nFz^{n-1}dz$.
Identifying $d{\cal S}$ and $dS$ we get
\be
-F_p=Res\,z^{-p}dS=Res\,z^{-p}d{\cal S}=-\sum a_j\sigma_{jp}-
\sum T_nq_{pn}
\label{EE}
\ee
which provides a formula for $F_p$ (note $\partial_na_j=0$ as indicated
in \cite{cc,ia}).  
\\[3mm]\indent
Next from Section 3 we produce ${\cal F}=(1/2)\psi\bar{\psi}+(X/i\epsilon)$
with a relation (\ref{AKQ}) between $F$ and ${\cal F}$.  Also a number
of formulas are given relating variables $\psi,\,\,\bar{\psi},\,\,
S=S^0,\,\,P=S_X=S_X^0,\,\,\phi=\bar{\psi}/2\psi=\partial{\cal F}/
\partial(\psi^2),\,\,\chi=|\psi|^2,$ and $\xi=(1/2\hbar)S$ in various
contexts.  In Section 4 the $dKdV_{\epsilon}$ theory is introduced
via $F=F^0$ in (\ref{five}), leading to $S=S^0+\epsilon S^1$ with 
$S^0$ (imaginary) in (\ref{OK}) and $|\psi|^2=exp(2\Re S^1)$ as in 
(\ref{OF}) ($S^1$ real).  The requirements of \cite{fa} produce
the constraint $|\psi|^2\Im\,P=-1$ for $P=S^0_X$ and one has fundamental
relations
\be
\phi=\frac{1}{2}exp\left(-\frac{2i}{\epsilon}\Im S^0\right)\,\,\,(cf.\,\,
(\ref{OM});\,\,X=-\epsilon\Im{\cal F};
\label{EF}
\ee
$$P=-\frac{i}{2\Re{\cal F}};\,\,log(2\phi)=-4i\xi\,\,\,(cf.\,\,(\ref{OOO})$$
plus the fundamental relation (\ref{OP}) for $dX\wedge dP$.  
In \cite{cb} a Hamilton
Jacobi theory for dispersionless theory was developed whose mission was
basically to motivate the treatment of $X$ in a canonical manner
commensurate with its role in \cite{fa}.  This is actually achieved at
the first WKB level $\psi=exp[(1/\epsilon)S+O(1)]$ but the $dKdV_{\epsilon}$
theory is needed e.g. to produce a meaningful expression for $|\psi|^2$.
\\[3mm]\indent
We now make some new computations to link various quantities.  First use
$V=-2F''$ as in (\ref{AKQ}) and recall (\ref{AA}); then (\ref{AA}) becomes
\be
\epsilon^2\psi''+2F''\psi=-E\psi
\label{EG}
\ee
Writing $\psi=exp[(1/\epsilon)S+S^1]$ this yields
\be
\epsilon^2(P^1_X+(P^1)^2)+\epsilon(P_X+2PP^1)+P^2+2F''=-E
\label{EH}
\ee
Equating powers of $\epsilon$ and recalling $P=iQ$ is imaginary with 
$P^1=S^1_X$ real one obtains
\be
F''=\frac{1}{2}(Q^2-E);\,\,P_X+2PP^1=0;\,\,P^1_X+(P^1)^2=0
\label{EI}
\ee
The second equation is consistent with remarks after (\ref{six}) (with $F^1=0$)
and the last equation then seems to determine $P^1$, which is an illusion
since more terms arise upon writing $\psi=exp[(1/\epsilon)S+S^1+\epsilon
S^2+\cdots]$.  Indeed using three terms we obtain
\be
P^2+2F''=-E;\,\,P_X+2PP^1=0;\,\,P_X^1+(P^1)^2+2PP^2=0
\label{EJ}
\ee
The first two equations are the same and the third shows that $P^1$
is not fixed by (\ref{EI}).  Since relations between the $P^i\sim P_i$ here 
must agree with relations based on (\ref{LL}) or (\ref{ZZZ}) we expect
(\ref{EJ}) (expanded with $F^2$ as in (\ref{EX})) to be compatible
with (\ref{FN}) for example.
However in order to deal with (\ref{ZZZ}) as a primary object in the
spirit of \cite{fa}, we have concentrated on balancing powers of
$\epsilon$ in (\ref{ZZZ}) based formulas (cf. however Remark 5.10
for balancing based on (\ref{EX})).
Then writing $8F''+4E=4Q^2=-4P^2$ we
obtain from (\ref{AKQ}) the equation ($F'''=-PP'=QQ'$)
\be
\epsilon^2{\cal F}'''+4Q^2\left({\cal F}'-\frac{1}{i\epsilon}\right)
+4QQ'\left({\cal F}-\frac{X}{i\epsilon}\right)=0
\label{EK}
\ee
which relates ${\cal F},\,\,P,$ and $X$.  One should check here the
consistency of (\ref{EK}) with (\ref{EF}) relating $X,\,\,P,\,\,\Re{\cal F},$
and $\Im{\cal F}$.  Thus $\Im{\cal F}=-(X/\epsilon)$ and $\Re{\cal F}=
-(i/2P)$ so ${\cal F}=-(1/2Q)-(iX/\epsilon)\,\,(P=iQ)$ and
(\ref{EK}) becomes (${\cal F}'-(1/i\epsilon)=Q'/2Q^2$ and ${\cal F}-
(X/i\epsilon)=-1/2Q$)
\be
\epsilon^2\left[\frac{Q'''Q-6Q''Q'+6(Q')^2}{2Q^3}\right]+4Q^2\frac{Q'}{2Q^2}
+4QQ'\left(\frac{-1}{2Q}\right)=0\Rightarrow
\label{EL}
\ee
$$\Rightarrow Q'''Q-6Q''Q'+6(Q')^3=0\equiv \left(\frac{1}{Q}\right)'''=0
\sim\frac{1}{Q}=AX^2+BX+C$$
\indent
{\bf HEURISTIC OBSERVATION 5.1.}$\,\,$ In the framework 
indicated after (\ref{six}),
the equation (\ref{EL}) essentially determines
$Q$, along with $P^1$ from (\ref{EI}) and $P^2$ from
(\ref{EJ}) ($P=iQ$).  
This says that the class of potentials $V\,\,(V=-2F'')$ admitting
an $(X,\psi)$ duality via ${\cal F}$ and arising from a KdV connected
WKB expansion is restricted to $Q$ satisfying (\ref{EL}), which
in turn essentially determines the entire $\epsilon$ expansion for 
$\tilde{P}=P+\epsilon P^1+\epsilon^2 P^2+\cdots$. The restriction on $Q$ is
however removed in the expanded theory to follow 
and Theorem 5.5 below provides clarification.  
Note that,
given in addition a Riemann surface background,
$A,\,B,\,C$ in 
(\ref{EL}) can depend on $T_n\,\,(N\geq 2)$ and $a_j$ (see also
the expanded development below).
\\[3mm]\indent {\bf REMARK 5.2.}$\,\,$ This theorem indicates 
some aspects of the kind
of relation between Riemann surfaces and $(X,\psi)$ duality which
was sketched heuristically in \cite{ca}.
\\[3mm]\indent
{\bf REMARK 5.3.}$\,\,$  It should be no surprise that a KdV connection might
restrict the WKB term $Q$ but we will see below that in fact there is no
such restriction on $Q$.  Evidently 
the $(X,\psi)$ duality will be
generally meaningful for $\psi=exp[(1/\epsilon)S^0+S^1]$ with $S^0$
imaginary and $\Re S^1\not= 0$.
\\[3mm]\indent
Next one asks for any possible relations between the $a_i$ and ${\cal F}$
for example and in fact such relations exist.  Given $F\sim F(a,T)$ as in 
(\ref{J}) one has a connection of the $a_i$ to ${\cal F}$ through $P$
via (\ref{EI}) or (\ref{EJ}) for exmple.  We recall also that $\partial_n
a_j=0$ and in the background there are Whitham equations of the form
\be
\frac{\partial d\omega_j}{\partial a_i}=\frac{\partial d\omega_i}{\partial 
a_j};\,\,\partial_nd\omega_j=\frac{\partial d\Omega_n}{\partial a_j};\,\,
\partial_nd\Omega_m=\partial_md\Omega_n
\label{EM}
\ee
(cf. \cite{cc,na}).  Given now that $d\omega_j=-\sum_1^{\infty}\sigma_{jm}
z^{m-1}dz$ along with the standard
$d\Omega_n=[-nz^{-n-1}-\sum_1^{\infty}q_{mn}z^{m-1}]dz$ the
equations (\ref{EM}) imply e.g. (cf. \cite{cn})
\be
\partial_pq_{mn}=\partial_nq_{mp};\,\,\partial_n\sigma_{jm}=\frac{\partial
q_{mn}}{\partial a_j};\,\,\frac{\partial \sigma_{jm}}{\partial
a_i}=\frac{\partial 
\sigma_{im}}{\partial a_j}
\label{EN}
\ee
In particular this indicates that $\partial_Xq_{mn}$ and $\partial_X\sigma_
{jm}$ make sense.  We could now compute $F''=F_{XX}$ from (\ref{J}) but it is
simpler to use (\ref{EE}) where
\be
F'=\sum_1^ga_j\sigma_{m1}+\sum_1^{\infty}T_nq_{1n}
\label{EQ}
\ee
from which
\be
F''=\sum_1^ga_j\sigma'_{j1}+q_{11}+\sum_1^{\infty}T_nq'_{1n}
\label{ER}
\ee
Such a formula shows that in fact $P$ is connected to the $a_j$ and hence
so is ${\cal F}$.  Thus in particular ${\cal F}=-(1/2Q)-(iX/\epsilon)$
with $2F''=Q^2-E\,\,(P=iQ)$ so
\be
Q^2=\frac{1}{4(\Re{\cal F})^2}=E+2\left[\sum_1^ga_j\sigma'_{j1}+q_{11}
+\sum_1^{\infty}T_nq'_{1n}\right]
\label{ES}
\ee
We summarize in (see Theorem 5.9 for a more proper theorem)
\\[3mm]\indent
{\bf NAIVE THEOREM 5.4.}$\,\,$ The prepotentials ${\cal F}$ and $F$ are related
to the $a_i$ via (\ref{J}) and (\ref{EQ}) - (\ref{ES}).  Further
\be
2Q\frac{\partial Q}{\partial a_k}=2\sigma'_{k1}=-\frac{1}{2}
\left(\frac{\partial \Re{\cal F}/\partial a_k}
{(\Re{\cal F})^3}\right)
\label{ET}
\ee
However $\partial\Re{\cal F}/\partial a_k=\partial {\cal F}/\partial a_k$
which implies
\be
\frac{\partial{\cal F}}{\partial a_k}=-\frac{1}{2}\frac{\partial log\,Q}
{\partial a_k}
\label{EU}
\ee
\indent
The development in Theorem 5.1 can be expanded 
as in Theorem 5.9 below when the framework for
$F$ is enlarged to $F=F_0+\epsilon F_1+\cdots$ (which we disallowed
after (\ref{six}) for convenience).  Thus if one assumes $F=F^0+\epsilon
F_1+\cdots$ for example then with $\psi=exp[(1/\epsilon)S+S_1
+\epsilon S_2+\cdots]$ (\ref{EG}) becomes at low order 
(writing
$\psi'=[(1/\epsilon)P+P_1+\epsilon P_2]\psi$ and
$\psi''=[(1/\epsilon)P'+P_1'+\epsilon P_2'+\{(1/\epsilon)P
+P_1+\epsilon P_2\}^2]\psi$ - we use $P_i$ or $P^i$ interchangeably)
$$
\epsilon^2\left[\frac{1}{\epsilon}P'+P_1'+\epsilon P_2'
+\frac{1}{\epsilon^2}P^2+P_1^2+\epsilon^2P_2^2+\frac{1}{\epsilon}
2PP_1+2PP_2+2\epsilon P_1P_2\right]+$$
\be
+2\left(F_0+\epsilon F_1+\epsilon^2F_2\right)''=-E
\label{EW}
\ee
leading to $P^2+2F_0''+E=0$ as before, plus (think of $P_{2i+1}$ as real
and $P_{2i}$ as imaginary)
\be
P'+2PP_1+2F_1''=0;\,\,P_1'+P_1^2+2PP_2+2F_2''=0;\,\,\cdots
\label{EX}
\ee
(so $F_{2i+1}$ is imaginary and $F_{2i}$ is real - we
will take $F_{2i+1}=0$ in order to have real potentials and use
arguments of \cite{ch}).
Thus $8F''+4E=\Upsilon=-4P^2-\epsilon(4P'+8PP_1)-\epsilon^2(P_1'+P_1^2
+2PP_2)+\cdots$ with ${\cal F}-(X/i\epsilon)=-(1/2\Im\,S')=-(1/2\Im\,
\tilde{P})$ where $\Im\,S'=2(Q+\epsilon\Im\,P_1+\epsilon^2\Im\,P_2+\cdots)
=\Im\tilde{P}$ and ${\cal F}'-(1/i\epsilon)=(\Im\,\tilde{P}'/2(\Im
\,\tilde{P})^2)$.  Then (\ref{AKQ}) becomes
\be
\epsilon^2{\cal F}'''+\frac{1}{2\Im\,\tilde{P}}\left[\frac{\Im\,\tilde{P}'}
{\Im\,\tilde{P}}\Upsilon-\frac{\Upsilon'}{2}\right]=0
\label{EZ}
\ee
Here $P=iQ$ and $P^2=-Q^2$ with 
\be
\Upsilon=4Q^2+\epsilon\Upsilon_1+\epsilon^2\Upsilon_2+\cdots;\,\,
\Upsilon'=8QQ'+\epsilon\Upsilon_1'+\epsilon^2\Upsilon_2'+\cdots
\label{FA}
\ee
with $\Upsilon_{2i+1}$ imaginary and $\Upsilon_{2i}$ real,
and $\Im\,\tilde{P}=Q+\epsilon{\cal P}$ 
(which should correspond to $Q+\sum_1^{\infty}\epsilon^{2i}\hat{P}_{2i}$
with ${\cal P}=\sum_1^{\infty}\epsilon^{2i-1}\hat{P}_{2i}$ and
$P_{2i}=i\hat{P}_{2i}$)
so that
\be
\frac{\Im\,\tilde{P}'}{\Im\,\tilde{P}}=\frac{Q'+\epsilon{\cal P}'}
{Q+\epsilon{\cal P}}=
\frac{Q'+\epsilon{\cal P}'}{Q}\left[1+\epsilon\frac{{\cal P}}{Q}
+\cdots\right]
\label{FB}
\ee
Hence the bracket $[\,\,\,\,]$ in (\ref{EZ}) has the form $(Q'/Q)
[4Q^2+\epsilon(\,\,\,)]-(1/2)[8QQ'+\epsilon (\,\,\,)_1]=
\epsilon
\{\,\,\,\,\}$ which leads to
\be
\epsilon^2{\cal F}'''+\frac{\epsilon}{2Q}\{\,\,\,\}=0;\,\,
{\cal F}'''=\frac{1}{2}\left[\frac{\Im\,\tilde{P}'}{(\Im\,\tilde{P})^2}
\right]''
\label{FC}
\ee
and the leading term from ${\cal F}'''$ will be the same as in (\ref{EL}).
Now the first terms
in (\ref{FC}) will involve (note ${\cal P}=\epsilon\hat{P}_2+\cdots$)
\be
{\cal F}'''=\left[\frac{1}{2}\frac{Q'+\epsilon{\cal P}'}{Q^2}\left(
1+\epsilon\frac{{\cal P}}{Q}+\cdots\right)^2\right]''=\frac{1}{2}\left[
\frac{Q'}{Q^2}+\epsilon\left(\frac{{\cal P}'}{Q^2}+\frac{Q'{\cal P}}{Q}
\right)+O(\epsilon^2)\right]''=
\label{FD}
\ee
$$=\frac{1}{2}\left(\frac{Q'}{Q^2}\right)''+\epsilon [\,\,\,]''+O(\epsilon^2)
=\frac{Q'''Q-6Q''Q'+6(Q')^3}{2Q^3}+\epsilon[\,\,\,]''+O(\epsilon^2)$$
and the first balance involves the $\epsilon$ term in $(\epsilon/2Q)
\{\,\,\,\}$ of (\ref{FC}) which can be extracted from (see below
for an expansion and note ${\cal P}=\epsilon\hat{P}_2+\cdots$)
\be
\frac{1}{2(Q+\epsilon{\cal P})}\left\{\left[\frac{Q'}{Q}+\epsilon
\left(\frac{Q'{\cal P}}{Q^2}+\frac{{\cal P}'}{Q}\right)\right]\cdot
(4Q^2+\epsilon\Upsilon_1)-\frac{1}{2}(8QQ'+\epsilon\Upsilon_1')\right\}=
\label{FE}
\ee
$$=\frac{1}{2(Q+\epsilon{\cal P})}\left\{\epsilon\frac{Q'}{Q}\Upsilon_1
-\epsilon\frac{\Upsilon'_1}{2}\right\}$$
But $\Upsilon_1=-4(P'+2PP_1)=0$ from (\ref{EX}) with $F_1=0$.  Hence
the $\epsilon$ term is automatically zero and there is no restriction
imposed here on $Q$.
We check now the next balance (which is at the same level
as (\ref{EL})).  Thus the $\epsilon^2$ term in (\ref{EZ}) will have
an $\epsilon^2$ term from (\ref{FE}) which should involve
$$
\Theta=\frac{1}{2(Q+\epsilon{\cal P})}\left\{\left[\frac{Q'}{Q}+
\epsilon\left(\frac{Q'{\cal P}}{Q^2}+\frac{{\cal P}'}{Q}\right)+
\epsilon^2\left(\frac{{\cal P}{\cal P}'}{Q^2}+\frac{Q'{\cal P}^2}{Q^2}
\right)\right](4Q^2+\epsilon\Upsilon_1+\epsilon^2\Upsilon_2)-
\right.$$
\be
\left.
-\frac{1}{2}(8QQ'+\epsilon\Upsilon_1'+\epsilon^2\Upsilon'_2)\right\}
\label{FI}
\ee
Setting ${\cal P}=\epsilon P_2+\epsilon^3P_4+\cdots$
and recalling
\be
\Upsilon_1=-4iQ'-8iQP_1;\,\,\Upsilon_2=-P_1'-P_1^2+2Q\hat{P}_2
\label{FJ}
\ee
we obtain
\be
\Theta\sim\frac{1}{2Q}\left(1+\epsilon^2\frac{\hat{P}_2}{Q}\right)\cdot\left\{
\left[\frac{Q'}{Q}+\epsilon^2\left(\frac{Q'\hat{P}_2}{Q^2}+
\frac{\hat{P}_2'}{Q}\right)
\right.\right.+
\label{FK}
\ee
$$\left.+\epsilon^4\left(\frac{\hat{P}_2\hat{P}_2'}{Q^2}+
\frac{Q'\hat{P}_2^2}{Q^2}\right)\right]
\left[4Q^2-\epsilon(4iQ'+8iQP_1)-\epsilon^2(P'_1+P_1^2-2Q\hat{P}_2)\right]-$$
$$\left.
-\frac{1}{2}\{8QQ'-\epsilon[4iQ''+8i(Q'P_1+QP_1')]-\epsilon^2[P_1''
+2P_1P_1'-2(Q'\hat{P}_2+Q\hat{P}_2')]\}\right\}$$
The $\epsilon^2$ term from $\Theta$ is then
\be
\frac{1}{2Q}\left\{
-\frac{Q'}{Q}(P_1'+P_1^2-2Q\hat{P}_2)\right.-
\label{FL}
\ee
$$-\left.4Q^2
\left(\frac{Q'\hat{P}_2}{Q^2}+\frac{\hat{P}_2'}{Q}\right)+\frac{1}{2}
[P_1''+2P_1P_1'-2(Q'\hat{P}_2+Q\hat{P}_2')]\right\}$$
so adding this to $\epsilon^2{\cal F}'''$ we require
\be
0=\frac{Q'''Q-6Q''Q'+6(Q')^3}{2Q^3}-
\label{FM}
\ee
$$-\frac{Q'}{2Q^2}(P_1'+P_1^2-2Q\hat{P}_2)-\frac{4}{2Q}(Q'\hat{P}_2
+\hat{P}'_2Q)+$$
$$+\frac{1}{4Q}[P_1''+2P_1P_1'-2(Q'\hat{P}_2+Q\hat{P}_2')]$$
Using again $Q'+2QP_1=0$ as a determination of $P_1$ with $\Upsilon_2
=-P_1'-P_1^2+2Q\hat{P}_2=2Q\hat{P}_2-(Q'/2Q)^2+(Q''/2Q)-(1/2)(Q'/Q)^2$
this yields
\be
0=\frac{Q'''Q-6Q''Q'+6(Q')^3}{2Q^3}+\frac{Q'}{2Q^2}\left[2Q\hat{P}_2+
\frac{Q''}{2Q}-\frac{3}{4}\left(\frac{Q'}{Q}\right)^2\right]-
\label{FN}
\ee
$$-\frac{2}{Q}(Q'\hat{P}_2+\hat{P}_2'Q)+\frac{1}{4Q}\left[2Q\hat{P}_2
+\frac{Q''}{2Q}-\frac{3}{4}\left(\frac{Q'}{Q}\right)^2\right]'$$
This can be then regarded as as a determination of $\hat{P}_2$ and we have
\\[3mm]\indent {\bf THEOREM 5.5.}$\,\,$ An expanded treatment of the
context of Heuristic Observation 5.1 shows that no restriction on $Q$
is required and the development will provide 
(modulo possible ``fitting" clarified below) a recursive
procedure determining the $P_i$, with first terms
$P_1=-Q'/2Q$ and $\hat{P}_2$ determined by (\ref{FN}).
\\[3mm]
\indent {\bf REMARK 5.6.}$\,\,$ Theorem 5.5 generates the $P_i$, hence
the $S_i$, and this must agree with what comes from $F=F_0+\epsilon F_1
+\cdots$.  Given $F_0$ related to KdV as above this would seem to generate
some $F_i$ via (\ref{five}),
but then a fitting
problem may arise with possibly hopeless constraints.  Thus we must
expand also the expressions based on (\ref{five}) where $F=F_0$
and consider a full $dKdV_{\epsilon}$ theory in some sense.  This is begun
after Remark 5.5 but the development should be coupled with a deeper
examination of the early terms.
\\[3mm]\indent
In order to expand Remark 5.6 we consider $F=\sum_0^{\infty}\epsilon^k
F^k$ and look at the early terms.  If we remain in the context of KP
or KdV then (\ref{five}) should be implemented with
\be
F\left(T_n-\frac{\epsilon}{n\lambda^n}\right)-F(T_n)=
\sum_{k=0}^{\infty}\epsilon^k\left[-\epsilon\sum_1^{\infty}\left(
\frac{F_n^k}{n\lambda^n}\right)+\frac{\epsilon^2}{2}\sum\frac{F_{mn}^k}
{mn}\lambda^{-m-n}+O(\epsilon^3)\right]
\label{FP}
\ee
Here one is specifying $\epsilon$ as the scale factor in $T_n=\epsilon
t_n$ etc. and it is common to the expansion of $F$ and the vertex
operator calculations.  This yields then from $log\psi=(1/\epsilon)
\sum T_n\lambda^n +(1/\epsilon)\sum_1^{\infty}S_{j+1}\lambda^{-j}\sim
(1/\epsilon^2)\Delta F+(1/\epsilon)\sum_1^{\infty} T_n\lambda^n$ with
$S\to\tilde{S}=\sum_1^{\infty}T_n\lambda^n+\sum_0^{\infty}\epsilon^k
\sum_1^{\infty}S^k_{j+1}\\
\lambda^{-j}=S^0+\sum_1^{\infty}\epsilon^k
\sum_1^{\infty}S^k_{j+1}\lambda^{-j}$ and one has
$$
\frac{1}{\epsilon}\sum_0^{\infty}\epsilon^k\left[-\sum_1^{\infty}\left(
\frac{F_n^k}{n}\right)\lambda^{-n}+\frac{\epsilon}{2}\sum\left(
\frac{F_{mn}^k}{nm}\right)\lambda^{-m-n}+O(\epsilon^2)\right]=$$
\be
=\frac{1}{\epsilon}\sum_0^{\infty}\epsilon^k\sum_1^{\infty}S^k_{j+1}\lambda^{-j}
\label{FQ}
\ee
(note here that lower indices correspond to derivatives and upper indices
are position markers except for $S_{j+1}^k$ where $j+1$ is a position
marker).
Hence in particular
\be
\sum_1^{\infty}S^0_{j+1}\lambda^{-j}=-\sum_1^{\infty}\left(\frac{F^0_n}
{n}\right)\lambda^{-n};
\label{FR}
\ee
$$\sum_1^{\infty}S^1_{j+1}\lambda^{-j}=-\sum_1^{\infty}\left(\frac
{F_n^1}{n}\right)\lambda^{-n}+\frac{1}{2}\sum_1^{\infty}\left(\frac
{F_{nm}^0}{mn}\right)\lambda^{-m-n}$$
leading to
\be
S_{j+1}^0=-\frac{F^0_j}{j}\,\,\,(as\,\,\,before);\,\,S_2^k=-F_1^k;\,\,
S_{j+1}^k=-\frac{F^k_j}{j}+\frac{1}{2}\sum_1^{j-1}
\left(\frac{F^{k-1}_{m,(j-m)}}{m(j-m)}\right)
\label{FS}
\ee
for $k\geq 1$, together with
\be
\tilde{P}=\tilde{S}_X=P+\sum_1^{\infty}\epsilon^k\sum_1^
{\infty}\partial_XS^k_{j+1}\lambda^{-j}
=P+\sum_1^{\infty}\epsilon^kS_X^k;\,\,S^k=\sum_1^{\infty}S^k_{j+1}\lambda^{-j}
\label{FT}
\ee
where $S^k_{j+1}$ is given in (\ref{FS}) and $P^k=P_k=\sum_1^{\infty}
\partial_XS^k_{j+1}\lambda^{-j}$.  Now following the patterns in Section 4
we want $S^0$ imaginary with 
\be
S^0_X=P=iQ=\lambda-\sum_1^{\infty}\left(\frac{F_{1j}}{j}\right)
\lambda^{-j}
\label{FU}
\ee
while $S^1$ and $P^1=\partial_XS^1$ should be real, etc.  
and similar expansions apply for $S^k,\,\,P^k$, etc. (cf. \ref{YR})).
\\[3mm]\indent
The spirit of KdV now gives $\partial_{2n}F^0=0$ and $F^0_{1,2n}=0$
etc. as in Section 4 (following \cite{ch}) and there seems to be no
reason why we cannot extend this to $F_{2n}=0$ and $F_{1,2n}=0$ via
$F^k_{2n}=0$ and $F^k_{1,2n}=0$, provided $F$ is real (cf. \cite{ch}).
Then as in Section 4, $P=\lambda-\sum_1^{\infty}[F^0_{1,2n-1}/(2n-1)]
\lambda^{1-2n}=iQ$, and e.g. (cf. (\ref{six}))
\be
P^1=\partial_XS^1=\partial_X\sum_1^{\infty}S^1_{j+1}\lambda^{-j}=
\label{FW}
\ee
$$=\partial_X\left[-\lambda^{-1}F_1^1+\sum_2^{\infty}\left(-\frac
{F_j^1}{j}\lambda^{-j}
+\frac{1}{2}\sum_{j\geq 2\,\,even}\lambda^{-j}\sum_1^{j-1}\left\{
\frac{F^0_{2m-1,j-2m+1}}{(2m-1)(j-2m+1)}\right\}\right)\right]$$
(the terms $F^0_{mn}$ vanish for $m$ or $n$ even so one has only
$F^0_{2m-1,2n-1}\lambda^{-2(m+n)+2}$ terms which can be labeled
as $\lambda^{-j}F^0_{2m-1,j-2m+1}$ for $j$ even).  Now $P^1$ real along
with $F$ real would be nice and (for $\lambda=ik$) 
a realization for this could
be begun via $F_j^1=0$ or simply $F^1=0$.
This situation also came up before in a pleasant way (cf. also (\ref{EX})) so
let us stipulate $F^{2i+1}=0$ and see what happens.  In particular this
drops the $F^1$ term from (\ref{FW}) and $P^1$ is then real as desired.
Further when we do this the lowest order terms involved in 
(\ref{OK}) - (\ref{OP}) remain the same but additional terms arise.
Thus consider $P\to \tilde{P}=P+\sum_1^{\infty}\epsilon^kP_k$ with $P_{2i}$
imaginary and $P_{2i+1}$ real so in (\ref{AAQ}) - (\ref{AGQ}) one replaces 
$P$ by $\tilde{P}$ and $S$ by $\tilde{S}=S^0+\sum_1^{\infty}\epsilon^k
S^k$ where we have concentrated positive powers of $\lambda$ in $S^0$.
From (\ref{FS}) we will have only $F^{2s}$ terms now which are real
and $S^k_{j+1}$ involves $F^k_j$ and $F^{k-1}_{m,(j-m)}$ so for $k=2n$ even
we have $S^{2n}_{j+1}=-F_j^{2n}/j$ while for $k=2n+1$ odd $S_{j+1}^
{2n+1}=(1/2)\sum_1^{j-1}[F^{2n}_{m,(j-m)}/m(j-m)]$ which can be rewritten
as in (\ref{FW}).  This says
\be
S^{2n} =\sum_1^{\infty}S^{2n}_{j+1}\lambda^{-j}=-\sum_1^{\infty}\left(
\frac{F_j^{2n}}{j}\right)\lambda^{-j}=-\sum_0^{\infty}\left(\frac{F^{2n}_
{2m+1}}{2m+1}\right)\lambda^{-2m-1};
\label{FX}
\ee
$$S^{2n+1}=\sum_1^{\infty}S_{j+1}^{2n+1}\lambda^{-j}=\frac{1}{2}\sum_
{j\geq 2\,\,even}\lambda^{-j}\sum_1^{j-1}\left(\frac{F^{2n}_{2m-1,j-2m+1}}
{(2m-1)(j-2m+1)}\right)$$
so $S^{2n}$ is imaginary and $S^{2n+1}$ is real for $\lambda=ik$.
Then in (\ref{AAQ}) - (\ref{AGQ}) and (\ref{ON}) - (\ref{OOO}) we have
$|\psi|^2\Im\tilde{P}=-1$ with e.g.
\be
log(2\phi)=-\frac{2i}{\epsilon}\Im\tilde{S}=\frac{2i}{\epsilon}
\left[S^0+\sum_1^{\infty}\epsilon^{2n}S^{2n}\right]
\label{FY}
\ee
Again one has $X=-\epsilon\Im{\cal F}$ so (for $P_{2n}=iQ_{2n}=i\hat{P}_{2n}$)
\be
{\cal F}=-\frac{1}{2\Im\tilde{P}}-\frac{iX}{\epsilon};\,\,
-\frac{1}{2\Re{\cal F}}=\Im\tilde{P}=Q+\sum_1^{\infty}\epsilon^{2n}Q_{2n}
\label{FZ}
\ee
Hence in place of the ``naive" Theorem 4.1 one would want to consider perhaps
\be
\frac{dX}{\epsilon}\wedge d\Im\tilde{P}=\frac{dX}{\epsilon}\wedge dQ+
dX\wedge \sum_1^{\infty}\epsilon^{2n-1}dQ_{2n}=
\left(\frac{1}{2(\Re{\cal F})^2}\right)d\Im{\cal F}\wedge d\Re{\cal F}
\label{GA}
\ee
(it seems appropriate to retain the scale factor $\epsilon$ with $X$ 
here).  Therefore as an expansion of Theorem 4.1 we have
\\[3mm]\indent {\bf THEOREM 5.7.}$\,\,$ In the expanded framework
just indicated one has (\ref{GA}).
\\[3mm]\indent
We note also that the potential $V$ now has the form $V=-2\partial^2_XF$
so with $F^{2k+1}=0$ and $F^{2k}$ real we have
\be
V=-2\sum_0^{\infty}\epsilon^{2k}F^{2k}_{XX}
\label{FV}
\ee
Such a formula could arise via $v=v(x,t_i)\to v(X/\epsilon,T_i/\epsilon)=
V(X,T)+O(\epsilon)$ in any case (as indicated after (\ref{BB}) and here
we are simply representing the $O(1/\epsilon)$ terms in $\tau=
exp[(1/\epsilon^2)F^0+O(1/\epsilon)]$ in an explicit form.
It does represent a restriction on possible potentials however.
\\[3mm]\indent
Now look at the expanded framework and retrace the argument (\ref{ET}) - (\ref
{FN}) to see whether our procedure is adapted to determine the $F^{2n}$
with $F^{2n+1}=0$, and what is involved.  
(we revise this procedure in Remark 5.10 and deal with an alternative 
balancing based on (\ref{EX}), (\ref{FX}), etc.).
If we take $F^1=0$ in (\ref{EX})
then $P'+2PP_1=0$ or $Q'+2QP_1=0$ and this was useful in balancing as well
as determining $P_1$ from $Q$.  Note that $Q=Q(k)$ via $Q^2=2F^0_{XX}-E$ 
where $\lambda^2=-E=-k^2$ and an expansion (\ref{YR}) holds. 
Thereafter the next balance is indicated
in (\ref{FN}) which serves to determine $P_2$ and therefrom $S^2$ and $F^2$
via 
\be
P_2=\partial_XS^2=-\sum_1^{\infty}\left(\frac{F^2_{1,2m+1}}{2m+1}\right)
\lambda^{-2m-1}
\label{GB}
\ee
Thus the $F^2_{1,2m+1}$ are in principle determined by residues from
$P_2$ and we defer momentarily the question of complete determination
of $F^2$.  The next balance arising from (\ref{EZ}) will involve the
$\epsilon$ term from ${\cal F}'''$ in (\ref{FD}) and the $\epsilon^3$
term in $(\epsilon/2Q)\{\,\,\,\}$ in (\ref{FC}).  Thus the $\epsilon$
term in ${\cal F}'''$ appears to be $(1/2)[({\cal P}'/Q^2)+(Q'{\cal P}/Q)]$ but
${\cal P}=\epsilon\hat{P}_2$ and hence there is no $\epsilon$ term.  For
the $\epsilon^3$ term in $(\epsilon/2Q)\{\,\,\,\}$ we go to (\ref{FC}) and 
write (recall $\Upsilon_1=0$)
\be
\frac{\epsilon}{2Q}\{\,\,\,\}=\frac{1}{2Q}\left[1+\epsilon
\left(\frac{{\cal P}}{Q}\right)+\cdots\right]\left\{\left[\frac{Q'}{Q}+
\epsilon\left(\frac{Q'{\cal P}}{Q^2}+\frac{{\cal P}'}{Q}\right)+\right.\right.
\label{GC}
\ee
$$+\left.\left.\epsilon^2\left(\frac{Q'}{Q}\left(\frac{{\cal P}}{Q}\right)^2
+\frac{{\cal P}'{\cal P}}{Q^2}\right)\right]\cdot (4Q^2+\epsilon^2\Upsilon_2)
-\frac{1}{2}(8QQ'+\epsilon^2\Upsilon'_2)\right\}$$
and one sees that there is no $\epsilon^3$ term (recall ${\cal P}\sim
\sum_1^{\infty}\epsilon^{2i-1}\hat{P}_{2i}$).  Thus the balancing act occurs
for even powers $\epsilon^{2n}$ only and will determine the $\hat{P}_{2n}$ in
terms of $Q$.  Then using (\ref{FX}) one can find $F^{2n}_{1,2m+1}$ by 
residues, and subsequently the $F^{2n}_{1,2m-1,j-2m+1}$ by differentiation,
leading to $P_{2n+1}$.  Hence
\\[3mm]\indent {\bf THEOREM 5.8.}$\,\,$   The procedure indicated is
consistent and in principle allows determination of the $P_n$ and $F^{2n}$ 
from $Q$.
\\[3mm]\indent {\bf THEOREM 5.9.}$\,\,$ In the Riemann surface context
the relation $2F''=Q^2-E$, with $F''$ given by (\ref{ER}),
determines $Q$ as a function of $T_n\,\,(n\geq 2)$ and $a_j$.  Hence
by Theorem 5.8 one knows $\tilde{P}$ as a function of $T_n$ and $a_j$.
Then since $\tilde{Q}=\Im\tilde{P}=-(1/2\Re{\cal F})$ we have in place of
(\ref{ET}) - (\ref{EU}) the formula
\be
\frac{\partial{\cal F}}{\partial a_k}=-\frac{1}{2}\frac{\partial\,log
\tilde{Q}}{\partial a_k}
\label{GE}
\ee
\indent
{\bf REMARK 5.10.}$\,\,$ The balancing via (\ref{ZZZ}) as in Theorem 5.5
leading to Theorem 5.8 can be accomplished in an alternative way, which has
some simpler aspects, by working with (\ref{EX}), (\ref{FX}), etc.
Indeed, extending the calculations (\ref{EJ}) with $F=\sum F^{2n}\epsilon^
{2n}$ one obtains
\be
-Q^2+2F^0_{XX}+E=0;\,\,2QP_1 +Q'=0;\,\,P_1'+P_1^2-2Q\hat{P}_2+2F^2_{XX}=0;
\label{GF}
\ee
$$\hat{P}_2'+2QP_3+2P_1\hat{P}_2=0;\,\,P_3'-\hat{P}_2^2-2Q\hat{P}_4
+2P_1P_3+2F^4_{XX}=0;\cdots$$
Putting in power series as in (\ref{FW}) and (\ref{FX}) one can equate
coefficients of powers of $\lambda=ik$.  For example one can consider
$F^2_{11}=Q\hat{P}_2-(1/2)(P_1'+P_1^2)$ along with the expansions
\be
i\hat{P}_2--\partial_X\sum_0^{\infty}\left(\frac{F^2_{2m+1}}{2m+1}\right)
(ik)^{-2m-1}\Rightarrow \hat{P}_2=F^2_{11}k^{-1}+\cdots;
\label{GG}
\ee
$$P_1=\frac{1}{2}\sum_1^{\infty}(-1)^pk^{-2p}\sum_1^{2p-1}
\left(\frac{F^0_{2m-1,2p-2m+1}}{(2m-1)(2p-2m+1)}\right)$$
from (\ref{FW}) - (\ref{FX}), and $Q$ given via (\ref{YR}).
We have not checked the details of calculation here.
\\[3mm]
\indent {\bf REMARK 5.11.}$\,\,$  In conclusion we can say that, given
a dKdV potential $V=-2F^0_{XX}$ arising from a finite zone KdV situation
(and leading to $Q$), one can create a $dKdV_{\epsilon}$ context in which
Theorems 5.7 - 5.9 are valid.  In the absence of a finite zone connection
one still has all formulas indicated except for those involving the
$a_j$.  Possible ``direct" connections to quantum mechanics can arise
as indicated in the beginning of Section 3.

\end{document}